\documentclass[onecolumn,11pt]{IEEEtran}
\usepackage{amssymb, amsmath, graphicx}

\newcommand{\R}{\mathbb{R}}
\newcommand{\Z}{\mathbb{Z}}

\newcommand{\QED}{\hfill $\Box$}
\newcommand{\limn}{\ensuremath{\underset{n \rightarrow \infty}{\lim\,}}}
\newenvironment{Proof}[1]{\noindent \textbf{Proof#1: }}{\QED}
\newtheorem{Theorem}{Theorem}
\newtheorem{Lemma}{Lemma}
\newtheorem{Corollary}{Corollary}

\title{Packet Speed and Cost\\
in Mobile Wireless Delay-Tolerant Networks} 

\author{ \IEEEauthorblockN{
      Riccardo Cavallari\IEEEauthorrefmark{2}, Stavros Toumpis\IEEEauthorrefmark{1},  Roberto Verdone\IEEEauthorrefmark{2}, and Ioannis Kontoyiannis\IEEEauthorrefmark{1}\IEEEauthorrefmark{3}}
			
\IEEEauthorblockA{\IEEEauthorrefmark{2}Department of Electrical, Electronic, and Information Engineering, University of Bologna, Italy} \\			
\IEEEauthorblockA{\IEEEauthorrefmark{1}Department of Informatics, Athens University of Economics and Business, Greece} \\
\IEEEauthorblockA{\IEEEauthorrefmark{3}Department of Engineering, University of Cambridge, UK  \\ 
Email: riccardo.cavallari@unibo.it, toumpis@aueb.gr, roberto.verdone@unibo.it, i.kontoyiannis@eng.cam.ac.uk}}

\begin{document}

\maketitle 

\begin{abstract}
A mobile wireless delay-tolerant network (DTN) model is proposed and analyzed, in which infinitely many nodes are initially placed on $\R^2$ according to a uniform Poisson point process (PPP) and subsequently travel, independently of each other, along trajectories comprised of line segments, changing travel direction at time instances that form a Poisson process, each time selecting a new travel direction from an arbitrary distribution; all nodes maintain constant speed. A single information packet is traveling towards a given direction using both wireless transmissions and sojourns on node buffers, according to a member of a broad class of possible routing rules. For this model,  we compute the long-term averages of the speed with which the packet travels towards its destination and the rate with which the wireless transmission cost accumulates. Because of the complexity of the problem, we employ two intuitive, simplifying approximations; simulations verify that the approximation error is typically small. Our results quantify the fundamental trade-off that exists in mobile wireless DTNs between the packet speed and the packet delivery cost. The framework developed here is both general and versatile, and can be used as a starting point for further investigation\footnote{Parts of this work appear, in preliminary form, in~\cite{cavallari1, cavallari2, cavallari3}. This work has been submitted to the IEEE Transactions on Information Theory.}.
\end{abstract}



\newpage

\section{Introduction}

In delay-tolerant networks (DTNs), packet delivery delays are often comparable to the time it typically takes for the network topology to change substantially. This means that packets have the opportunity to take advantage of such topology changes. An important class of DTNs is that of mobile wireless DTNs, where communication is over a wireless channel and changes in the topology are due to node mobility. Such networks appear in disparate settings, and may be comprised of sensors, smartphones, vehicles, and even satellites~\cite{dtn_book}.

We propose and analyze a mobile wireless DTN model consisting of an infinite number of nodes moving on an infinite plane. Each node moves with constant speed along a straight line, choosing a new travel direction (from a given distribution) at time instances forming a Poisson process. Nodes move independently of each other. It is assumed that a single information packet needs to travel to a destination located at an infinite distance in a given direction; two modes of travel are possible: wireless transmissions and physical transports on the buffers of nodes. Wireless transmissions are instantaneous, but come at a transmission cost that is a function of the vector specifying the change in the packet location due to the transmission (and not simply the distance covered, therefore the cost may be anisotropic). Physical transports, on the other hand, do not have an associated cost, but they introduce delays. The packet alternates between the two modes of travel using a routing rule selected from a broad class of such rules that are described in terms of two quantities: a forwarding region and a potential function. 

In this setting, we define two performance metrics that characterize the specific routing rule used. The first one is the packet speed, defined as the limit, as the packet travel time goes to infinity, of the ratio of the distance covered divided by the packet travel time. The second one is the normalized packet cost, defined as the limit, as the packet travel time goes to infinity, of the ratio of the cost incurred divided by the distance covered.  

Because of the generality and mathematical complexity of this model, in order to compute the values of these two metrics we introduce two simplifying approximations that allow us to use tools from the theory of Markov chains: the {\em Second-Order Approximation} of Section \ref{subsection111},  and the {\em Time-Invariance Approximation} of Section \ref{subsection77jj}. Both approximations judiciously introduce renewals in the node mobility process. Under these assumptions, we show that the packet's travel direction can be described as a discrete-time, continuous-state Markov chain, where each time slot of the chain corresponds either to a packet transmission or to a time interval during which the packet travels along a straight line on the buffer of a node. 

Under general, natural assumptions on the class of routing rules considered (see Sections~\ref{routingruleclass},~\ref{section6hg}, and~\ref{section1t7}), we show that the Markov chain is uniformly (geometrically) ergodic, its transition kernel can be precisely identified, and we can compute its unique invariant distribution. The actual values of the two performance metrics can then be computed explicitly in terms of this distribution.

Simulations results (see Figs~\ref{experiment1}, \ref{experiment2} and~\ref{experiment3} in Section~\ref{s:results}) show that the errors introduced by the approximation are modest and typically small; namely, less that 10\% on the average for the numerical results we present. Furthermore, the qualitative trends and trade-offs revealed by our analytical results are in all cases confirmed by the simulation experiments. In particular, it is demonstrated that, as expected, the packet can travel faster towards its destination, but only at a higher transmission  cost due to the more frequent use of wireless transmissions, and vice-versa. 

The rest of this paper is organized as follows. In Section~\ref{section_related} we discuss related work in this field. In Section~\ref{section_model} we introduce the precise network model and the corresponding performance metrics. The analysis is carried out in Sections~\ref{section113}, \ref{section17}, and \ref{section1t7}. In Section~\ref{section_results} we present numerical results for a specific setting. Section~\ref{section_conclusions} contains some concluding remarks. Finally, in the Appendix we present some of the more technical proofs and computations. 

\newpage

\section{Relation to Prior Work}
\label{section_related}

Since Gupta and Kumar's celebrated work \cite{gupta1} on networks with immobile nodes, their asymptotic analysis has been adapted by various authors to the study of networks of mobile nodes employing both wireless transmissions and physical transports. Notably, in a line of work initiated in \cite{tse1} and continued by, among others, \cite{diggavi2005even}, \cite{toumpis9}, and \cite{sharma3}, trade-offs between  throughput and delay were explored. In these works, routing protocols that make use of the direction in which each node is traveling were not considered. Such protocols were examined in \cite{jacquet10} for a network of finite area, where mobile nodes move along straight lines and change travel direction at random times forming a Poisson process. There, it was assumed that each node creates packets for an immobile destination whose location is known to them, and nodes employ  the \emph{Constrained Relative Bearing} geographic routing protocol: each packet remains on the buffer of a node when that node is moving effectively enough (i.e.,  along a sufficiently good direction towards its destination), otherwise the packet is transmitted to a more suitable node whenever such a node is available nearby. This scheme was shown to achieve near-constant throughput per node with bounded delivery delay, asymptotically as the number of nodes in the network increases. Compared to these works that pursue network-wide analysis, we take a more `local' view, focusing on the long-term cost and delay in the forwarding of a single packet, and avoiding the calculation of relevant metrics up to multiplicative constants. 

Recently, the topic of percolation in mobile wireless networks, i.e., the replication of a single packet across the network through a combination of wireless transmissions and physical transports, has attracted significant interest. In such settings, the packet percolates with finite speed, except in the trivial case when the node density is sufficiently high so that a giant network component exists at any time instant. The problem of computing this speed has been studied, e.g., in~\cite{jacquet17}, where the authors consider two- or higher-dimensional networks with a wide range of mobility models, and in \cite{li14}, where the replication of the packet is constrained. In the present work we study the travel of a single packet copy towards a specific destination, as opposed to its spread by replication over the whole network; this significantly differentiates both the application of our work and the flavor of our analysis. 

Numerous works have focused on one-dimensional mobile wireless DTN models, with nodes constrained to move along a common, fixed line. Such models are suitable for vehicular DTNs of cars moving on highways and are motivated in part by questions related to road safety issues. For example, the authors of~\cite{baccelli78} consider a highway comprising two lanes of vehicles moving in opposite (westbound and eastbound) directions; all vehicles travel with the same speed, and the distances between cars in each lane are exponentially distributed (with different means for the two lanes). There, in their study of the speed with which a single packet travels in the eastbound direction using both modes of transport, the authors identify two distinct regions in the propagation of the packet: depending on the specific values of the problem parameters, either the packet is essentially moving with the speed of the nodes, or its speed increases quasi-exponentially with the node densities. More general versions of such models are studied in~\cite{zarei12} and \cite{baccelli79}. Compared to these works, our two-dimensional model is significantly more challenging. Moreover, by properly selecting the distribution of the node travel directions, our results can be applied to urban settings where these directions are appropriately constrained. 

Although all the works mentioned so far are of a mainly theoretical nature, there have been a number of works with simulation studies of hybrid routing protocols that employ both modes of transport. Numerous different protocols have been considered; for example, the Moving Vector (MoVe) protocol \cite{lebrun1} favors transmissions to nodes that are scheduled to pass \emph{the closest} from the destination,  whereas the AeroRP protocol~\cite{peters1} favors nodes that are traveling \emph{the fastest} towards the destination; see \cite{tasiopoulos2} and references therein for other such examples. Compared to these works, our analysis gives theoretical results on the performance of a general class of routing protocols.

Tools of stochastic geometry have also been employed in studying networks where node mobility is crucial to their performance but, in contrast to all prior work mentioned above, there is \emph{no physical transport} of data. For example, the authors of \cite{madadi1} investigate a model in which a mobile node moves along a straight line on a plane where stationary base stations (BSs) are placed according to a Poisson point process; the node is in contact with a BS if the two are closer than a threshold distance. In this setting, the authors show that the node comes in contact with the BSs according to an alternating-renewal process; this observation can be used for studying the quality of service (QoS) experienced by the node if it streams video through the BSs and for computing the distribution of download times of files downloaded by the node through the BSs. In \cite{liu9} the authors study a wireless sensor network comprising mobile sensors distributed on an infinite plane; each sensor moves along a straight line in a fixed random direction and at a random speed, sensing for `targets.' For this model, the authors compute the values of various performance metrics related to the quality of target coverage provided by the network; namely, they compute the percentage of the area covered at any given time instant as well as the time needed for a target located outside the coverage region to be sensed, for both mobile and immobile targets.  Finally, in~\cite{peres1} the authors study a network of nodes moving according to independent Brownian motions in $\R^d$; two nodes are in contact whenever they are within some threshold distance from each other. Here the authors study three important random quantities: the time until a target (mobile or immobile) comes in contact with any of the nodes, the time until the nodes come in contact with all points in a given subset, and the time until a target comes in contact with a node belonging to a giant network component. Although all these works do not involve the physical transport of information, the tools we develop in this work may be applicable to many of the scenarios they consider. For example, the incidence rates derived in Section~\ref{section17} can be used in a mobile sensor setting to compute the rate with which a mobile sensor with an arbitrary sensing region encounters targets.

As already mentioned, the mathematical complexity of our two-dimensional network model and the generality of the routing protocols we consider have necessitated the use of approximations. An alternative approach is to avoid approximations altogether, arriving at exact results, but starting with a much simpler network model. This is the approach taken in \cite{cheliotis1}, where a one-dimensional discrete-time network is studied. There, the network consists of $n$ locations, arranged on a ring, on which two mobile nodes perform independent random walks. A single packet travels in the clockwise direction on the buffer of one of the nodes, and it only gets transmitted from one node to the other when these are collocated, the current packet carrier is moving in the counter-clockwise direction, and the other node is traveling in the clockwise direction. Using probabilistic tools from the theory of Markov chains, explicit expressions are derived for the long-term average packet speed and for the steady-state average number of wireless transmissions per time slot.

Finally, we note that elements of the analysis at hand first appeared in \cite{sidera2, sidera4}. Compared to the work at hand, the work there notably differs as follows: \emph{(i)} Regarding the network model used, nodes do not change their travel direction and a more restrictive class of routing rules is used. \emph{(ii)} Regarding the developed analysis, packet trajectories are modeled using an i.i.d. process (as opposed to a Markov chain) and an approximation cruder than the Second-Order Approximation is used (we elaborate on the difference between the two approximations in Section~\ref{subsection111}). 

\newpage

\section{Network Model}
\label{section_model}

\subsection{Node mobility}
\label{sec78}
At time $t=0$, infinitely many nodes are placed on the plane $\R^2$ according to a uniform Poisson point process (PPP) with \textbf{(node) density} $\lambda>0$. Subsequently, each node travels on $\R^2$, independently of the rest of the nodes, according to the following random waypoint mobility model (here and in the rest of this work, travel directions are specified in terms of the angle $\theta\in[-\pi,\pi)$ they form with the positive direction of the $x$-axis): the node selects a random travel direction $D_1$ according to a (not necessarily uniform) \textbf{direction density} $f_D:[-\pi,\pi) \rightarrow [0,\infty)$; the node moves in this direction along a straight line with constant \textbf{node speed} 
$v_0>0$, for a random duration of time $E_1$ that follows an exponential distribution with mean $1/r_0$; the parameter $r_0>0$ is the \textbf{(node) turning rate}. The node then picks another random travel direction $D_2$ from the same density $f_D(\cdot)$, and travels in that direction for another exponentially distributed amount of time $E_2$ (again with mean $1/r_0$), and so on, ad infinitum. The random variables (RVs) $\{D_i\}$ and $\{E_i\}$ are all independent of each other. 

The density $f_D(\cdot)$ can be used to describe situations where the nodes have preferred travel directions; for example, in a Manhattan-like city center, we expect most nodes to be traveling along two main axes.

For our results to hold, we require that there is an $\epsilon_D>0$ such that $f_D(\cdot)$ does not take values in $(0,\epsilon_D)$, i.e., $f_D(\cdot)$ can be $0$ but in the set where it is not $0$ it is bounded away from it. However, in order to keep the exposition simple, in the rest of this discussion we also assume that $f_D(\cdot)$ is strictly positive everywhere. Indeed, if $f_D(\cdot)$ is zero on some subset of $[-\pi,\pi)$, then this set can be removed from consideration and all subsequent analysis applies without change. 

Note that the time instants when the travel direction of a given node changes form a Poisson process with rate $r_0$, and by the displacement theorem \cite[Theorem 1.10]{baccelli3}, at any time instant $t>0$, the locations of all nodes follow a PPP with density $\lambda$. Also note that, because the distribution of the duration of time a node keeps its travel direction is not a function of its current direction, the travel direction of a given node at any fixed time $t>0$ has density $f_D(\cdot)$.

\subsection{Transceiver model}

Nodes are equipped with transceivers with which they can exchange packets. Suppose node~1 wants to transmit a packet to node~2, whose relative location with respect to node~1 is described by the vector $\mathbf{r}=(r,\phi)$; that is, node~2 is at a distance $r=|\mathbf{r}|\geq 0$ from node~1, in the direction $\phi\in[-\pi,\pi)$. We assume that such a transmission has \textbf{(wireless transmission) cost} $C(\mathbf{r})$, for some fixed cost function $C(\cdot)$. We also assume that all packets have the same length, and that all transmissions are instantaneous.

Some remarks on our transceiver model choice are in order. First, the cost $C(\mathbf{r})$ can be used to model the energy dissipated by the transmitter in order for the packet to reach a relative location $\mathbf{r}$~\cite{ephremides3}, or the cost (in lost throughput) of having to silence other transmitters so that the transmission is received correctly by the receiver~\cite{gupta1}. Second, allowing the cost to be a function of the vector $\mathbf{r}$ and not just its length $r=|\mathbf{r}|$ allows us to treat cases where there is anisotropy in the environment; for example, in a Manhattan-like environment we expect the energy dissipated for transmitting at a given distance in the directions of the street/avenues 
to be less than the energy in other directions, as the signals in the former case do not have to pass through as many buildings. 

Third, $C(\mathbf{r})$ can be interpreted as the \emph{expected value} of the transmission cost in case this is random, e.g., due to fading. All our results, appropriately interpreted, continue to hold in that case, provided the sources of randomness in the cost are independent from all other sources of randomness. We make no more reference to such interpretations in the rest of this work. Fourth, the assumption that the transmission is instantaneous is made for mathematical convenience, and it is very reasonable in our delay-tolerant context. Indeed, we are interested in measuring delays that are comparable to the time needed for the topology to change significantly, whereas the time needed for the transmission of a packet is typically such that the locations of the transmitter, the receiver and the other nodes in their vicinity do not change perceptibly.

Finally, our model does not explicitly capture the interaction between packets, i.e., there is no contention for the channel; the need for all packets to share the available bandwidth is implicitly modeled through the use of the wireless transmission cost function $C(\mathbf{r})$. 

\subsection{Traffic model}
\label{subsection7721}

We consider a single, \emph{tagged} packet, created at time $t=0$, that must travel to a destination placed at an infinite distance away from the packet source. With no loss of generality, we take the destination to be in the direction of the positive $x$-axis. 

The assumption that the packet destination is located at an infinite distance away is made for mathematical convenience; we plan to calculate performance metrics using the invariant distribution of a Markov chain, and for this reason it is necessary for the length of the packet travel to be infinite; we expect these metrics to be relevant in the design of real networks provided packets travel for finite but not small distances.

Given that the destination of the packet is in the direction of the positive $x$-axis, in the following, we define a travel direction $\theta_1$ to be \textbf{better} than a travel direction $\theta_2$ if $|\theta_1| <|\theta_2|$; therefore, if the packet changes its travel direction to a better one, given that all nodes travel with the same speed, it starts approaching its destination faster. We will also use the terms \textbf{equal}, \textbf{best}, \textbf{worse}, and \textbf{worst}, for travel directions, in the same sense.

The packet can travel to the destination using a hybrid geographic/delay-tolerant \textbf{routing rule} (RR) that uses combinations of wireless transmissions (the geographic part of the RR) and sojourns along the buffers of nodes (the delay-tolerant part of the RR). 

\subsection{Stages}
\label{subsection2}

Irrespective of the RR used, we can always break the travel of the tagged packet towards its destination into an infinite sequence of \textbf{stages} $i=1,2,\dots$, with each stage $i$ corresponding to either a single wireless transmission (in which case we call it a \textbf{(wireless) transmission stage} between the \textbf{transmitter} and the \textbf{receiver} of that stage), or a single sojourn on the buffer of a node, the {\bf carrier}, while its travel direction does not change (which we call a \textbf{buffering stage}). Observe that a new stage will occur even if the carrier changes its direction but the packet stays with it. Therefore, each stage is associated with exactly one of the linear segments comprising the packet trajectory. Since nodes change directions after exponential times and the packet destination is located at an infinite distance away from its source, there will be an infinite number of stages with probability 1.

With each stage $i=1,2,\dots,$ we associate a number of RVs. Firstly, let $\Theta_i \in [-\pi,\pi)$ denote the carrier travel direction in the case of buffering stages, and the travel direction of the \emph{receiver} in the case of transmission stages. Let $T_{i-1},T_i$  be the time instants when stage $i$ starts and ends, respectively, and $\Delta_i=T_i-T_{i-1}$ be its duration. Observe that $\Delta_i=0$ for transmission stages and $\Delta_i>0$ for buffering stages. 
Let $(X_{W,i},Y_{W,i})$ be the changes in the coordinates of the packet due to the wireless transmission at stage $i$, and let $C_i=C((X_{W,i},Y_{W,i}))$ be the associated transmission cost  so that, if $i$ is a buffering stage, then $X_{W,i}=Y_{W_i}=C_{W,i}=0$. Likewise, let $X_{B,i}$ be the change in the $x$-coordinate of the packet due to the buffering in stage $i$ so that, if that stage is a transmission stage, then $X_{B,i}=0$. Observe that $X_{B,i}=v_0 \Delta_i \cos \Theta_i$. Finally, write $X_i=X_{W,i}+X_{B,i}$. We will refer to any change of the $x$-coordinate as \textbf{progress}. We collect all these RVs in Table~\ref{table2}. 

\begin{table}[ht!]
\begin{center}
\caption{RVs associated with stage $i$} \label{table2}
\begin{tabular}{|c|c|} \hline
Quantity & Symbol \\ \hline 
Stage index & $i=1,2,\dots$ \\
Carrier (for buffering stage) or receiver (for transmission stage) travel direction during stage $i$ & $\Theta_i \in [-\pi,\pi)$ \\
Time instant stage $i$ starts & $T_{i-1}$ \\ 
Time instant stage $i$ ends & $T_i$ \\
Stage $i$ duration & $\Delta_i=T_i-T_{i-1}$ \\ 
Progress due to transmission in stage $i$ & $X_{W,i}$\\
$y$-coordinate change due to transmission in stage $i$ & $Y_{W,i}$\\
Cost of transmission in stage $i$ & $C_i=C \left( (X_{W,i},Y_{W,i})\right)$\\
Progress due to buffering during stage $i$ & $X_{B,i}=v_0 \Delta_i \cos \Theta_i$ \\
Progress during stage $i$ & $X_i=X_{B,i}+X_{W,i}$\\ \hline
\end{tabular}
\end{center}

\end{table}

\subsection{Performance metrics}
\label{subsection1}
We describe the performance of the RR employed in terms of the \textbf{(packet) speed} $V_p$, defined as,
\begin{equation}
V_p \triangleq \limn \frac{\sum_{i=1}^n X_i}{T_n}= 
\limn \frac{\sum_{i=1}^n X_i}{\sum_{i=1}^n  \Delta_i },
\label{eq1}
\end{equation}
and the \textbf{(normalized packet) cost} $C_p$,
\begin{equation}
C_p \triangleq \limn \frac{\sum_{i=1}^n C_i}{\sum_{i=1}^n X_i}.
\label{eq2}
\end{equation}
In the sequel we will show that, under appropriate conditions, these limits indeed exist and are constant, with probability~1.

The packet speed $V_p$ represents the limit of the average rate (in units of distance over time) with which the packet makes progress towards its destination, as the number of stages goes to infinity. Similarly, $C_p$ represents the limit of the average rate (in units of cost over distance) with which cost is accumulated in the long run as the packet progresses towards its destination.

Although it is straightforward to estimate the values of $V_p$ and $C_p$ through simulation, it is hard to determine them analytically. One reason is that the sequence $\{(X_{W,i},Y_{W,i},X_{B,i},\Delta_i)\;;\;i\geq 1\}$ does not form a Markov chain. Therefore, one would have to consider the complete continuous-time chain on an infinite-dimensional state
space describing the positions and travel directions of all nodes on the plane at any given time $t$; clearly this is a daunting task. For this reason, we introduce two approximation assumptions that create artificial regeneration epochs in the analysis. These assumptions are chosen in a judicious manner, allowing us both to apply tools from Markov chains, and to guarantee that the induced approximation errors in the computations of $V_p$ and $C_p$ are modest in size. This is indeed shown to be the case through numerous simulation examples, for a wide range of parameters.

Finally, we expect a trade-off to exist between the cost and the speed: if an efficiently designed RR makes heavy use of wireless transmissions, we expect the packet to travel fast towards its destination, but at a significant cost; on the other hand, if an efficiently designed RR makes light use of transmissions, the cost will be low but the packet will also make slow progress towards its destination. Our simulation results also verify the existence of this trade-off for the class of RRs considered in this paper.

\subsection{Routing rule}
\label{routingruleclass}

For the rest of this work we limit our attention to the following class of RRs, described in terms of a forwarding region and a potential function. First we need to introduce a simple notational convention.

\medskip

\noindent
{\bf Notation. } All node locations $\mathbf{r}$ in $\R^2$ are described in polar coordinates, $\mathbf{r}=(r,\phi)$ and they are always understood to be {\em relative} locations of one node relative to another, or relative to the origin $\mathbf{0}\in\R$. With a slight abuse of notation, we perform operations between locations as if they were
written in Euclidean coordinates. For example, if the locations of nodes~$A$ and~$B$ with respect to the origin are $\mathbf{r}_A$ and $\mathbf{r}_B$, respectively, then the location of~$B$ relative to~$A$ is $\mathbf{r}_B-\mathbf{r}_A$.

\medskip

Let the \textbf{Forwarding Region (FR)} $\mathcal{F}$ be the (nonempty) closed, bounded and convex subset of $\R^2$ defined as
\begin{equation*}
\mathcal{F} 
\triangleq \{\mathbf{r} \triangleq (r,\phi): 
-\pi \leq \phi < \pi ,~0 \leq r \leq b(\phi)\},
\end{equation*}
in terms of an arbitrary bounded  \textbf{boundary function} $b:[-\pi,\pi)\to[0,\infty)$; observe that $(0,0)=\mathbf{0} \in \mathcal{F}$.  We also assume throughout that the cost function $C(\cdot)$ is bounded on the bounded region ${\cal F}$. The FR of an arbitrary node $A$ located at $\mathbf{r}_A$ is
\begin{equation*}
\mathcal{F}(A)\triangleq 
\mathcal{F}\;\mbox{translated so that $\mathbf{0}$ is at $\mathbf{r}_A$}
\,=\, \mathbf{r}_A+\mathcal{F}.
\end{equation*} 

Suppose the packet is with a node $A$ at the origin. The suitability of a node within $\mathcal{F}(A)$ (either the current holder or another one) located at position $\mathbf{r}\in\mathcal{F}(A)$ and traveling in direction $\theta\in[-\pi,\pi)$ is described by the \textbf{potential} function $U(\theta,\mathbf{r})$; the higher the potential, the more suitable the node is. Different choices of the two functions $b$ and $U(\cdot,\cdot)$ give rise to different RRs within the class.  We make the following assumptions:

\medskip

\noindent
\textbf{Assumption 1. } $U(\cdot,\cdot)$ is a continuous, strictly monotonic function 
of $\theta$, in the following sense: if $|\theta_1|>|\theta_2|$, 
then $U(\theta_1,\mathbf{r}) <  U(\theta_2,\mathbf{r})$,
for any $\mathbf{r}$.

\medskip

\noindent
\textbf{Assumption 2. } If $U(\theta_1,\mathbf{r}_1)
<U(\theta_2,\mathbf{r}_2)$, then also 
$U(\theta_1,\mathbf{r}_1-\mathbf{r}_3)
<U(\theta_2,\mathbf{r}_2-\mathbf{r}_3)$, 
for any $\mathbf{r}_3$ such that both 
$\mathbf{r}_1-\mathbf{r}_3$ and 
$\mathbf{r}_1-\mathbf{r}_3$ belong to $\mathcal{F}$.

\medskip

Assumption 1 says that, if a node changes its travel direction to a strictly better one, then it becomes strictly more appealing for buffering the packet. Clearly, for any reasonable choice of the potential we should have that, if $|\theta_1|>|\theta_2|$, then $U(\theta_1,\mathbf{r}) \leq  U(\theta_2,\mathbf{r})$. Excluding the case of equality,  $U(\theta_1,\mathbf{r}) =  U(\theta_2,\mathbf{r})$, simplifies the analysis because it allows us to conclude that, at any time instant, all nodes in the same FR have different potentials, with probability~$1$. Allowing equality would require a longer but not substantially different analysis. The performance of protocols using potential functions where equality may hold can be approximated well by slightly modifying the potential, e.g., by adding a small corrective term $-\epsilon |\theta|$, for some $\epsilon>0$; therefore, this assumption does not limit significantly the scope of our work. 

Assumption 2 means that, if a node~$A$ located at $\mathbf{r}_1$ and traveling in direction $\theta_1$, is less appealing than a node~$B$ located at $\mathbf{r}_2$ and traveling with direction $\theta_2$, according to a node $C$ located at the origin, then node $A$ should also be less appealing than $B$ to any other node $D$ that has both $A$ and $B$ in its forwarding region. In other words, nodes should agree among themselves, at all times, about which of two nodes is better for buffering the packet; otherwise, there may be routing loops. Clearly, in this geographic routing 
context, any reasonable choice for the potential function should naturally satisfy this assumption.

Two more assumptions will be introduced later on in the analysis. Collectively, the four assumptions are satisfied for many, perhaps most, reasonable choices of the functions $b(\phi)$ and $U(\theta,\mathbf{r})$, adequately covering the spectrum of routing protocol design requirements. The assumptions are made partly for mathematical convenience, and they could be relaxed in various different directions without making the analysis substantially harder. We stress that our analysis does not require the specification of particular choices for the functions $b$ and $U(\cdot,\cdot)$, that is, of a particular RR; we consider a specific example in Section~\ref{section_results} where we present numerical results.

Having defined the all the key concepts, we can now specify the routing rule: 

\medskip

\noindent
{\bf Routing rule. }
The packet travels on the buffer of a carrier node $A_i$ until another node $A_{i+1}$, which we refer to as the \textbf{eligible} node, is found; $A_{i+1}$ is eligible if it lies in $\mathcal{F}(A_i)$ and its potential is greater than that of $A_i$ and of all other nodes within $\mathcal{F}(A_i)$. The packet is instantaneously transmitted to $A_{i+1}$ and the same rule is applied again. Then either another 
eligible node, $A_{i+2}$, is immediately found, in which case the packet is transmitted to $A_{i+2}$ at the same time instant, or a sojourn on the buffer of node 
$A_{i+1}$ is initiated; and so on.

\medskip

In Table~\ref{table1} we collect all the quantities used so far in modeling the network.

\begin{table}[ht!]
\begin{center}
\caption{Quantities and notation used in the network model specified 
in Section~\ref{section_model}}
\label{table1}
\begin{tabular}{|c|c|} \hline
Quantity & Symbol \\ \hline 
Node density & $\lambda$ \\
Direction density & $f_D(x), ~ x \in [-\pi,\pi)$ \\ 
Node speed & $v_0$ \\
Node turning rate & $r_0$ \\ 
Transmission cost & $C(\mathbf{r})$, $\mathbf{r} \in \R^2$ \\
Forwarding Region & $\mathcal{F}$ \\
Boundary function & $b(\phi)$, $~\phi \in [-\pi,\pi)$ \\ 
Potential & $U(\theta,\mathbf{r})$, $\theta \in [-\pi,\pi)$, $~\mathbf{r} \in \mathcal{F}$ \\ \hline
\end{tabular}
\end{center}
\end{table}

\subsection{Second-Order Approximation and its consequences}
\label{subsection111}

Here we introduce the first of our two approximations, which pertains to what happens between stages. 

\medskip

\noindent
\textbf{Second-Order Approximation:} \emph{
\begin{enumerate}
\item The moment a receiver $A$ receives the packet from a transmitter $B$, the complete mobility process is re-initialized, except that the position and travel direction of node $A$ are maintained and all nodes that appear (after the re-initialization) in $\mathcal{F}(A) \cap \mathcal{F}(B)$ and whose potential is greater than that of $A$ are removed. 
\item The moment a node $A$ carrying the packet changes its travel direction $\theta$ to a $\theta'$, the mobility process is re-initialized, except that $A$ maintains its position and travel direction and all created nodes within $\mathcal{F}(A)$ whose potential is greater than $\max\{U(\theta,\mathbf{0}),U(\theta',\mathbf{0})\}$ are removed.
\end{enumerate}}

Note that by `re-initialization' we mean that all nodes are placed in all of $\R^2$ again as they were at time $t=0$. Intuitively, the approximation introduces regeneration points in the mobility process, so that a Markov chain that is amenable to analysis may later on be defined. However, it does so without \emph{eligible} nodes unexpectedly appearing out of nowhere in the FR due to the re-initialization; \emph{ineligible} nodes do appear, however such nodes might have already been present in the FR before the re-initialization, so the re-initialization has the effect of reshuffling them, and the performance of the RR is not significantly affected. 

We call this approximation `second-order' to differentiate it from:
\begin{enumerate}
\item the \emph{First-Order Approximation} used in~\cite{sidera4} and \cite{cavallari1} (termed, there, \emph{Approximation 1}), under which, whenever a node $A$ receives the packet or changes its travel direction, the mobility process is regenerated, keeping node $A$'s position and travel direction, but without removing any nodes, and
\item the even coarser \emph{Basic Assumption} of \cite{sidera2} under which, whenever a node $A$ receives the packet, the mobility process is re-initialized keeping node $A$'s position but not its travel direction, and also without removing any nodes.
\end{enumerate}

We note that the derivations in~\cite{sidera4,cavallari1,sidera2}, which are based on these alternative approximations, are notably simpler, as more information is lost at each re-initialization and, in each setting, the trajectory of the tagged packet can be modeled with a random process simpler than that we eventually develop in Section~\ref{section1t7}.

\newpage

\section{Transmission Stage Analysis}
\label{section113}

\begin{figure}[ht!]
\begin{center}
\includegraphics{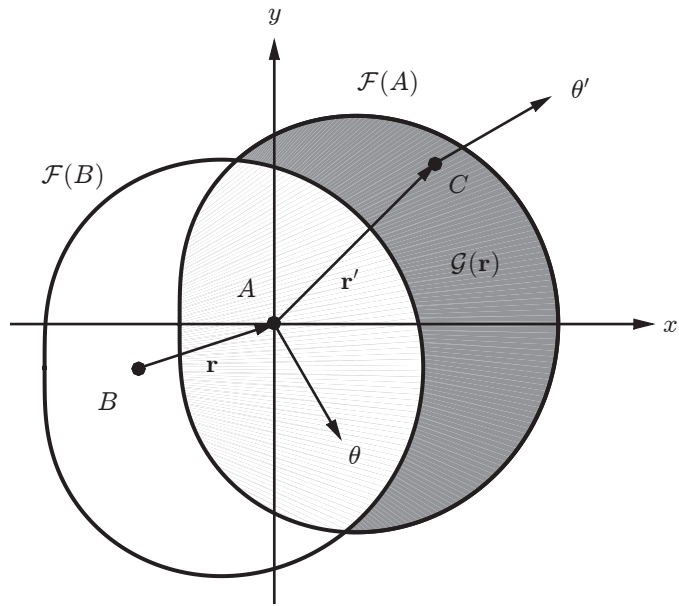}
\end{center}
\caption{The setting of Section~\ref{section113}. Here, there is an 
eligible node $C$ within $\mathcal{G}(\mathbf{r})$, however this 
is not always the case.}
\label{fig2}
\end{figure}

As the first step of the analysis, in this section we compute explicit expressions for a number of quantities related to what follows a wireless transmission stage.  The setting here, shown in Fig.~\ref{fig2}, is as follows: a node $A$ is traveling in direction $\theta \in [-\pi,\pi)$ and has just received the packet from some node $B$ such that the position of $A$ relative to $B$ is $\mathbf{r} \in \mathcal{F}(B)$. Our quantities of interest here are functions of $\theta$ and $\mathbf{r}$. Write
\begin{equation*}
\mathcal{G}(\mathbf{r}) \triangleq \mathcal{F}(A)\cap \mathcal{F}(B)^c,
\end{equation*}
for the locations in ${\cal F}(A)$ but not in ${\cal F}(B)$.

Let $E(N;\theta,\mathbf{r})$ be the expected number of nodes in $\mathcal{G}(\mathbf{r})$ whose potential is greater than that of $A$, so that
\begin{eqnarray}
E(N;\theta,\mathbf{r})&=& \int_{-\pi}^\pi \iint_{\mathcal{G}(\mathbf{r})} \lambda f_D(\theta') \mathbf{1} [U(\theta',\mathbf{r}')>U(\theta,\mathbf{0})]\, dA' d\theta'\nonumber  \\
&=& \int_{-\pi}^\pi \iint_{\mathcal{F}}\lambda f_D(\theta') \mathbf{1} [U(\theta',\mathbf{r}')>U(\theta,\mathbf{0}),~\mathbf{r}' \in \mathcal{G}(\mathbf{r})]\, dA'd\theta', \label{eq38}
\end{eqnarray}
where $dA'$ is the infinitesimal area element corresponding to $\mathbf{r}'$, and the indicator function $\mathbf{1}[\mathcal{X}]$ is equal to $1$ if the condition $\mathcal{X}$ holds and to $0$ if it does not.

Also, let $P_E(\theta,\mathbf{r})$ denote the probability of the event $\mathcal{E}(\theta,\mathbf{r})$ that $\mathcal{G}(\mathbf{r})$ does not contain an eligible node, i.e., that a new buffering stage will commence at the moment node $A$ receives the packet. This event will occur if there are no nodes in $\mathcal{G}(r)$ whose potential is greater than the potential of $A$. The number $N$ of such nodes has a Poisson distribution with mean $E(N;\theta,\mathbf{r})$, therefore,
\begin{equation}
P_E(\theta,\mathbf{r})=P(\mathcal{E}(\theta,\mathbf{r}))=\exp \left[ -E(N;\theta,\mathbf{r})\right].
\label{eq40402}
\end{equation}

Finally, let $g(\theta',\mathbf{r}';\theta,\mathbf{r})$ be the joint density of the location $\mathbf{r}' \in \mathcal{F}(A)$ and the travel direction $\theta' \in [-\pi,\pi)$ of the eligible node $C$ to which the packet  
is immediately transmitted from $A$ (see Fig.~\ref{fig2}), so that $g(\theta',\mathbf{r}';\theta,\mathbf{r})$ 
is equal to zero if no such node can exist for the given choice of $\theta'$ and $\mathbf{r}'$. In order to obtain a useful expression for  $g(\theta',\mathbf{r}';\theta,\mathbf{r})$ for all  $\mathbf{r}, \mathbf{r}' \in \mathcal{F} (A)$, $\theta,\theta' \in [-\pi,\pi)$, first 
observe that $g(\theta',\mathbf{r}';\theta,\mathbf{r})=0$ if $U(\theta,\mathbf{0}) \geq U(\theta',\mathbf{r}')$, i.e., node $A$ is at least as suitable as node $C$ for keeping the packet. We also have $g(\theta',\mathbf{r}';\theta,\mathbf{r})=0$ if $\mathbf{r}' \not \in \mathcal{G}(\mathbf{r})$, i.e., $\mathbf{r}'$ is in the intersection of the FRs $\mathcal{F}(A)$ and $\mathcal{F}(B)$ and so no eligible node may be found there by the Second-Order Approximation.

When both $U(\theta,\mathbf{0})<U(\theta',\mathbf{r}')$ and $\mathbf{r}' \in \mathcal{G}(\mathbf{r})$, the joint density of the location $\mathbf{r}'$ and direction $\theta'$ of node $C$ is $\lambda f_D(\theta')$, and $C$ will receive the packet if there is no other node in $\mathcal{G}(\mathbf{r})$ that is better than $C$. The expected number of such nodes is (cf. with the derivation of (\ref{eq38}))
\begin{multline*}
\int_{-\pi}^\pi \iint_{\mathcal{G}(\mathbf{r})}\lambda f_D(\theta'')\mathbf{1} [U(\theta'',\mathbf{r}'')>U(\theta',\mathbf{r}')]\, dA'' d\theta''  
\\= \int_{-\pi}^\pi \iint_{\mathcal{F}}\lambda f_D(\theta'') \mathbf{1} [U(\theta'',\mathbf{r}'')>U(\theta',\mathbf{r}'),\mathbf{r}'' \in \mathcal{G}(\mathbf{r})]\, dA''d\theta'', 
\end{multline*}
where, as before, $dA''$ is the infinitesimal area element corresponding to $\mathbf{r}''$, and as their number is Poisson distributed, we have
\begin{equation*}
g(\theta',\mathbf{r}';\theta,\mathbf{r})= \lambda f_D(\theta') 
\exp \left\{ -\int_{-\pi}^\pi \iint_{\mathcal{F}}\lambda f_D(\theta'') \mathbf{1} [U(\theta'',\mathbf{r}'')>U(\theta',\mathbf{r}'),\mathbf{r}'' \in \mathcal{G}(\mathbf{r})]\, dA''d\theta''\right\}.
\end{equation*}
Combining all cases, we have 
\begin{multline}
g(\theta',\mathbf{r}';\theta,\mathbf{r})= \lambda f_D(\theta')\mathbf{1}[U(\theta,\mathbf{0})<U(\theta',\mathbf{r}'),~\mathbf{r}' \in \mathcal{G}(\mathbf{r})] \\ \times \exp \left\{ -\lambda\int_{-\pi}^\pi \iint_{\mathcal{F}} f_D(\theta'') \mathbf{1} [U(\theta'',\mathbf{r}'')>U(\theta',\mathbf{r}'),\mathbf{r}'' \in \mathcal{G}(\mathbf{r})]\, dA''d\theta''\right\}.
\label{eq451d7}
\end{multline}

Observe that, for all $\theta \in [-\pi,\pi)$, $\mathbf{r} \in \mathcal{F}$, we must have 
\begin{equation}
P_E(\theta,\mathbf{r})+\int_{-\pi}^{\pi} \iint_{\mathcal{F}} g(\theta',\mathbf{r}';\theta,\mathbf{r})\, dA' d\theta' =1.
\label{eq4ho}
\end{equation}
This is due to the fact that, upon the reception of a packet, either a sojourn will start, or another transmission will take place, with probability $1$. 

\newpage

\section{Buffering Stage Analysis}
\label{section17}

As the second step of the analysis, in this section we compute explicit expressions for a number of quantities related to what follows a buffering stage. Specifically, suppose that at time $t=T_{i-1}$ a buffering stage $i$ starts with the packet in the buffer of a node $A$, and traveling in direction $\Theta_i=\theta \in [-\pi,\pi)$. The buffering ends at time $T_i=T_{i-1}+\Delta_i$, for some $\Delta_i>0$. 

We partition the event corresponding to the end of the buffering stage $i$ into four families of disjoint events, each one describing a different manner in which the buffering will end. We then use our second approximation, introduced in Section \ref{subsection77jj}, to compute the probability of each of these events.

\subsection{Four families of events}
\label{section6hg}
Given the value of $\Theta_i=\theta$, first we define the collection of events
\begin{equation*}
\mathcal{A}(\theta)=\{
\mathcal{A}(\theta,\theta')\;;\;
\theta' \in [-\pi,\pi)\},
\end{equation*}
where $\mathcal{A}(\theta, \theta')$ is the event that the buffering ends because, at time $T_i$, node $A$ changes its travel direction from $\theta$ to $\theta'$, while no eligible node is found. Second, we let
\begin{equation*}
\mathcal{B}(\theta)=\{
	\mathcal{B}(\theta,\theta',\mathbf{r}')\;;\;
	\theta'\in [-\pi,\pi),~\mathbf{r}'
	\in\mathcal{F}(A)
\}, 
\end{equation*}
where $\mathcal{B}(\theta,\theta',\mathbf{r}')$ is the event that the buffering ends because, at time $T_i$, node $A$ changes its travel direction from $\theta$ to some $\theta''$ and an eligible node is immediately found in location $\mathbf{r}'\in\mathcal{F}(A)$ and traveling in direction $\theta'$.

The third collection of events we will consider is
\begin{equation*}
\mathcal{C}(\theta)=
\{
\mathcal{C}(\theta,\theta',\mathbf{r}') \;;\;
\theta' \in [-\pi,\pi),~\mathbf{r}' \in \mathcal{F}(A)\},
\end{equation*}
where $\mathcal{C}(\theta,\theta',\mathbf{r}')$ is the event that the 
buffering ends because, at time $T_i$, while $A$ is still
traveling in direction $\theta$, a node $B$
located at $\mathbf{r}'\in\mathcal{F}(A)$
changes direction from some previous
$\theta''$ to $\theta'$, thus becoming eligible.

\begin{figure}[ht!]
\begin{center}
\includegraphics{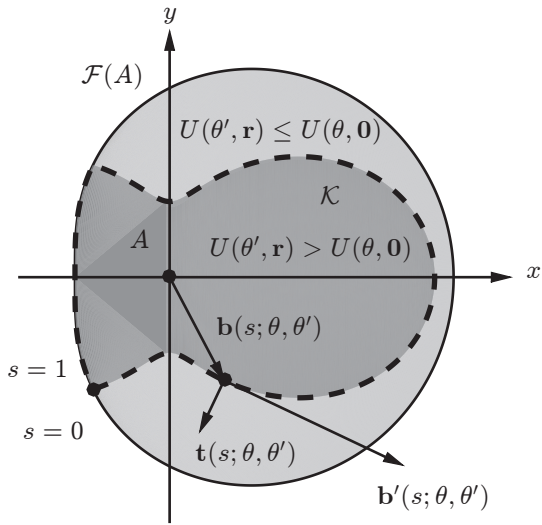}
\end{center}
\caption{Setting used in defining the family $\mathcal{D}(\theta)$.}
\label{fig78}
\end{figure}

To define the fourth family, we first need to introduce
another mild assumption on the FR $\mathcal{F}$ and the 
potential $U(\cdot,\cdot)$, complementing the assumptions of
Section~\ref{subsection7721} regarding the RR.
For any $\theta,\theta'$,
let $\mathcal{K}=\mathcal{K}(\theta,\theta')$ 
denote the subset of the FR of a node $A$ traveling
in direction $\theta$, where
$U(\theta',\mathbf{r})>U(\theta,\mathbf{0})$;
cf.~Fig.~\ref{fig78}. Therefore, 
nodes that enter $\mathcal{K}$ from the
outside immediately become eligible. 

\medskip

\noindent
{\bf Assumption 3. } We assume that, for any $\theta, \theta'\in[-\pi,\pi)$, 
the region $\mathcal{K}=\mathcal{K}(\theta,\theta')$ is convex.
Let the \textbf{threshold curve},
$\mathbf{b}(s; \theta,\theta')$, parametrized by $s \in [0,1]$,
be the curve separating $\mathcal{K}$ 
and $\mathcal{K}^c$. We assume that the
curvature of $\mathbf{b}$ is uniformly bounded, in that
$\mathbf{b}(s; \theta,\theta')$ is differentiable with
respect to $s$, for almost every $s\in[0,1]$, and there
exists a finite constant $M_b$, independent
of $\theta,\theta'$, such that the magnitude
of the derivative $\mathbf{b}'(s;\theta,\theta')$ 
with respect to $s$ is bounded by $M_b$:
$$|\mathbf{b}'|(s;\theta,\theta') \leq M_b,
\;\;\mbox{for almost all}\;s\in [0,1].$$
 
\medskip

Note that Assumption~3 implies that the length of the curve 
$\mathbf{b}(s; \theta,\theta')$ is bounded, a property
which is obviously satisfied for any reasonable choice of
the potential $U(\theta,\mathbf{r})$, provided the 
parametrization $\mathbf{b}(s; \theta,\theta')$ is suitably chosen.
[For concreteness, we also mention that 
the derivative $\mathbf{b}'$ with respect to $s$ above
is taken on the $x$- and $y$-coordinates
of $\mathbf{b}$.]
Let $\mathbf{t}(s;\theta,\theta')$ denote the unit vector 
perpendicular to the curve $\mathbf{b}(s; \theta,\theta')$ at the 
location specified by $s$, and pointing in the direction 
of lower potential. Observe that changing 
$s$ to $s+ds$ traces an infinitesimal line segment 
of length $ds |\mathbf{b}'|(s;\theta,\theta')$ 
that is perpendicular 
to $\mathbf{t}(s;\theta,\theta')$;
see Fig.~\ref{fig78}. 
Clearly, a node that ``hits'' the curve $\mathbf{b}$
from outside $\mathcal{K}$ immediately becomes eligible.

We can now define the last collection of
events of interest here:
\begin{equation*}
\mathcal{D}(\theta)=\{\mathcal{D}(\theta,\theta',s)\;;\; \theta' 
\in [-\pi,\pi),~s \in [0,1]\},
\end{equation*}
where $\mathcal{D}(\theta,\theta',s)$ denotes the event that the buffering 
ends because, at time $T_i$, an eligible node appears
at the position of the boundary of $\mathcal{K}$ corresponding
to $s$, traveling with direction $\theta'$.
Finally, we write
\begin{equation*}
\mathcal{U}(\theta)=\mathcal{A}(\theta) \cup \mathcal{B}(\theta) 
\cup \mathcal{C}(\theta) \cup \mathcal{D}(\theta),
\end{equation*}
and we note that $P\left(\cup_{E \in \mathcal{U}(\theta)} E |\Theta_i=\theta\right)=1$, i.e., these four cases cover every possible scenario, with probability~1.

\subsection{Transition rates}
\label{s:transition}

Let $\theta,\theta'\in[-\pi,\pi)$, $\mathbf{r}\in\mathcal{F}$ and $s\in[0,1]$
arbitrary,
let $dA'$ denote the infinitesimal area element 
in location $\mathbf{r}'$ as before,
and let $t=0$. With a slight abuse of notation we 
define the \textbf{transition rates} $r_\mathcal{A}(\theta,\theta')$, 
$r_\mathcal{B}(\theta,\theta',\mathbf{r}')$,  
$r_\mathcal{C}(\theta,\theta',\mathbf{r}')$, 
and $r_\mathcal{D}(\theta,\theta',s)$, as:
\begin{eqnarray}
r_\mathcal{A}(\theta,\theta')d\theta' dt&=&P(\mathcal{A}(\theta,\theta'), ~ \Delta_i=t| \Theta_i=\theta),\label{eq:RA}\\
r_\mathcal{B}(\theta,\theta',\mathbf{r}')d\theta' dA' dt&=& P(\mathcal{B}(\theta,\theta',\mathbf{r}'),~ \Delta_i=t|\Theta_i=\theta),\label{eq:RB}\\
r_\mathcal{C}(\theta,\theta',\mathbf{r}')d\theta' dA' dt&=& P(\mathcal{C}(\theta,\theta',\mathbf{r}'),~ \Delta_i=t|\Theta_i=\theta),\label{eq:RC}\\
r_\mathcal{D}(\theta,\theta',s)d\theta'  ds dt&=& P(\mathcal{D}(\theta,\theta',s),~ \Delta_i=t|\Theta_i=\theta).\label{eq:RD}
\end{eqnarray}
Intuitively, these rates describe the infinitesimal probability that a buffering stage will end exactly in each one of the four possible scenarios discussed above, after an infinitesimal duration of time $\Delta_i\in[0,dt]$. Formally, we could define $r_\mathcal{A}(\theta,\theta')$, via the limit
$$
P\left(\big[\cup_{\theta''\in(\theta'-\delta\theta'/2,
\theta'+\delta\theta'/2)}\mathcal{A}(\theta,\theta'')\big] \cap
\{\Delta_i \in[0,\delta t]\}
\Big|\Theta_i=\theta\right)
=
r_\mathcal{A}(\theta,\theta')\delta\theta'\delta t
+o(\delta\theta'\delta t),$$
as $\delta\theta',\delta t\downarrow 0$,
and similarly for the other three transition rates.
We now proceed to derive expressions for each of them,
in terms of the network model and the RR parameters
specified earlier; cf.~Table~\ref{table1}.
Again, with a slight abuse of terminology and
notation, in the subsequent discussion 
we omit the adjective ``infinitesimal'' most of
the time, e.g., referring to the quantities in
the right-hand sides of~(\ref{eq:RA})--(\ref{eq:RD})
simply as ``probabilities.'' 

Regarding $r_\mathcal{A}(\theta,\theta')$,  the probability 
in the right-hand side of (\ref{eq:RA}) is equal to the product 
of five different quantities:
$(a)$~the probability $r_0dt$ that node $A$ will change its direction during that interval;
$(b)$~the probability $f_D(\theta')d\theta'$ that $A$ will pick direction 
$\theta'$;
$(c)$ the probability that there are no eligible nodes in $\mathcal{F}(A)$ 
with potential at most $U(\theta,\mathbf{0})$ but greater than $U(\theta',\mathbf{0})$, 
which is (recall the derivation of (\ref{eq40402})),
\begin{equation*}
\exp \left\{ -\int_{-\pi}^{\pi} \iint_{\mathcal{F}} \lambda f_D(\theta'') 
\mathbf{1}[U(\theta,\mathbf{0})\geq U(\theta'',\mathbf{r}'')>U(\theta',\mathbf{0})]\, 
dA''d\theta'' \right\};
\end{equation*}
$(d)$~the probability, $p_1$, say, that no event in $\mathcal{C}(\theta)$ 
will occur before $\mathcal{A}(\theta,\theta')$;
and $(e)$~the probability, $p_2$, say, that that no event 
in $\mathcal{D}(\theta)$ will occur before $\mathcal{A}(\theta,\theta')$.

Now observe that $p_1$ is bounded below by
the probability $1-\lambda|\mathcal{F}(A)|r_0dt$ 
that no node in a region of area $|\mathcal{F}(A)|$ will change travel 
direction in a time interval of duration $dt$. 
As for $p_2$, we claim that it is bounded below by $1-2v_0\lambda M_b dt$,
where $M_b$ is the bound to the length of the 
curves $\mathbf{b}(s; \theta,\theta')$ specified 
by Assumption~3. Indeed, the expected 
number of nodes with a given travel direction $\theta'$ and density 
$\lambda f(\theta')d\theta'$ that cross $\mathbf{b}(s; \theta,\theta')$, 
whose length is less than $M_b$, in a time interval $[0,dt]$, with 
a relative speed less than $2v_0$, is less 
than $\lambda f(\theta') d\theta' M_b  2v_0dt $. Integrating
over all $\theta'$, it follows that the expected number of all such
nodes 
is less than $2v_0\lambda M_b dt$. As the distribution of their 
total number is Poisson, the probability that no node will cross some 
curve $\mathbf{b}(s; \theta,\theta')$ in a time interval $[0,dt]$ 
is greater than $1-2v_0\lambda M_b dt$, and so $p_2\geq 1-2v_0\lambda M_b dt$. We note that similar arguments can be used 
in the calculation of the other three transition rates to show that the 
probability that an event of a different type occurs does not affect 
the rate; as these arguments are straightforward, they will 
be omitted. 

Combining the above estimates
and ignoring terms of order $(dt)^2$, it follows that
\begin{equation}
r_\mathcal{A}(\theta,\theta')
=r_0 f_D(\theta')
\exp \left[-\int_{-\pi}^{\pi} \iint_{\mathcal{F}}\lambda f_D(\theta'') \mathbf{1}[U(\theta,\mathbf{0}) \geq U(\theta'',\mathbf{r}'')>U(\theta',\mathbf{0})]\, dA''d\theta''\right].
\label{rate_a}
\end{equation}

Regarding $r_\mathcal{B}(\theta,\theta',\mathbf{r}')$, note that,
if $U(\theta',\mathbf{r}')>U(\theta,\mathbf{0})$, then the probability  
in the right-hand side of (\ref{eq:RB}) is zero
because the condition implies that there was an eligible node 
before $A$ changed direction. However, 
if $U(\theta,\mathbf{0})\geq U(\theta',\mathbf{r}')$, then
this probability can again be expressed as the product of four different
terms: $(a)$~the probability $r_0 dt$ that node $A$ will change its travel 
direction during the interval $[0,dt]$; $(b)$~the probability
\begin{equation*}
\int_{-\pi}^{\pi} f_D(\theta'')\mathbf{1}\left[U(\theta',\mathbf{r}')>U(\theta'',\mathbf{0})\right]\, d\theta'',
\end{equation*}
that its new direction $\theta''$ will lead to a lower potential
than $U(\theta',\mathbf{r}')$ (otherwise, the packet would have stayed 
with $A$); $(c)$~the probability $\lambda f_D(\theta') dA' d\theta'$ that 
there is a node at the specified location $\mathbf{r}'$ with the specified 
travel direction $\theta'$; and $(d)$~the probability that there is no node 
in $\mathcal{F}(A)$ that is better than that node, which is 
(cf. with the derivation of (\ref{eq40402}))
\begin{equation*}
\exp \left[ -\int_{-\pi}^{\pi} \iint_{\mathcal{F}} \lambda f_D(\theta''') \mathbf{1}[U(\theta,\mathbf{0})\geq U(\theta''',\mathbf{r}''')>U(\theta',\mathbf{r}')]\, dA'''d\theta''' \right].
\end{equation*}
Therefore,
\begin{multline}
r_\mathcal{B}(\theta,\theta',\mathbf{r}') = r_0 \lambda  f_D(\theta') \mathbf{1}[U(\theta,\mathbf{0}) \geq U(\theta',\mathbf{r}')] \times\int_{-\pi}^{\pi} f_D(\theta'')\mathbf{1}\left[U(\theta',\mathbf{r}')>U(\theta'',\mathbf{0})\right]\,d\theta'' \\ \times \exp \left[ -\int_{-\pi}^{\pi} \iint_{\mathcal{F}} \lambda f_D(\theta''') \mathbf{1}[U(\theta,\mathbf{0}) \geq U(\theta''',\mathbf{r}''')>U(\theta',\mathbf{r}')]\, dA'''d\theta''' \right]. 
\label{rate_b}
\end{multline}

Regarding $r_\mathcal{C}(\theta,\theta',\mathbf{r}')$, the 
probability in the right-hand side of (\ref{eq:RC})
is zero when $U(\theta',\mathbf{r}')\leq U(\theta,\mathbf{0})$. 
Otherwise, it is equal to the probability that there is a node 
within the specified area, 
\begin{equation*} 
\lambda dA' \int_{-\pi}^{\pi} f_D(\theta'') \mathbf{1}[U(\theta'',\mathbf{r}')<U(\theta,\mathbf{0})]\, d\theta'',
\end{equation*}
multiplied with the probability 
$r_0 f_D(\theta') d\theta'dt$
that that node will turn to direction $\theta'$.
Therefore,
\begin{equation} 
r_{\mathcal{C}}(\theta,\theta',\mathbf{r}')  = \lambda  r_0 f_D(\theta') \mathbf{1}\left[U(\theta',\mathbf{r}')>U(\theta,\mathbf{0}) \right] \left[\int_{-\pi}^{\pi}f_D(\theta'') \mathbf{1}[U(\theta'',\mathbf{r}')<U(\theta,\mathbf{0})]\, d\theta'' \right].
\label{rate_c}
\end{equation}

\begin{figure}[ht!]
\begin{center}
\includegraphics{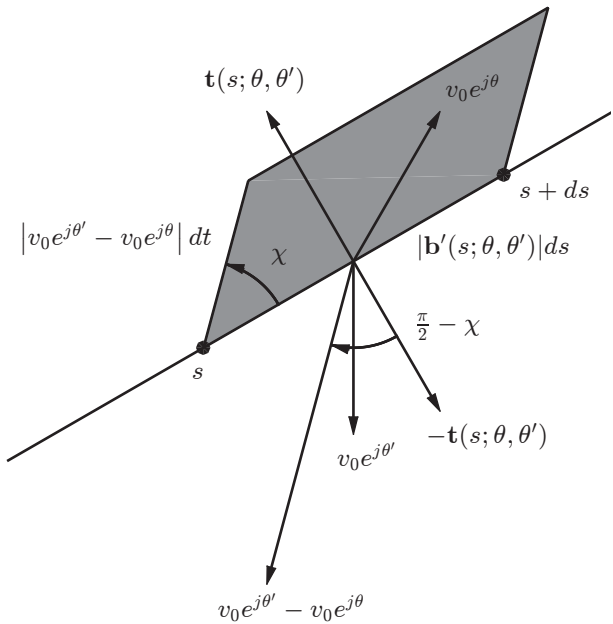}
\end{center}
\caption{The setting used in calculating the transition
rates $r_\mathcal{D}(s,\theta')$.}
\label{fig111}
\end{figure}

Regarding rate $r_\mathcal{D}(\theta,\theta',s)$, observe that nodes 
that move in direction $\theta'$ appear to node $A$ to be moving 
with relative speed $v_0 e^{j\theta'}-v_0 e^{j\theta}$;
cf.\ Fig.~\ref{fig111}. 
Also observe that, in order for the probability on the 
right-hand side of (\ref{eq:RD}) to be nonzero, the inner product 
$(v_0e^{j\theta'}-v_0e^{j\theta})\cdot \mathbf{t}(s;\theta,\theta')$ must 
be negative (as shown in the figure) so that nodes 
with travel direction $\theta'$ are hitting the boundary 
from the outside of $\mathcal{K}$. Then, this probability
is equal to the density of nodes $\lambda f_D(\theta') d\theta'$ 
traveling in direction $\theta'$, multiplied with the area of 
the parallelogram (which appears shaded in the figure) with sides of length $|v_0e^{j\theta'}-v_0e^{j\theta}|dt$ and $|\mathbf{b}'(s;\theta,\theta')|ds$, at an angle $\chi$.
Therefore, 
$$r_\mathcal{D}(\theta,\theta',s) 
=\mathbf{1}[(v_0e^{j\theta'}-v_0e^{j\theta})\cdot 
\mathbf{t}(s;\theta,\theta')<0] \lambda f_D(\theta') 
|v_0e^{j\theta'}-v_0e^{j\theta}|\,|\mathbf{b}'(s;\theta,\theta')|  \sin \chi,$$
and noting that the inner product,
\begin{equation*}
(e^{j\theta'}-e^{j\theta})\cdot (-\mathbf{t}(s;\theta,\theta'))=|e^{j\theta'}-e^{j\theta}| \cos(\pi/2-\chi)=|e^{j\theta'}-e^{j\theta}| \sin \chi,
\end{equation*}
we obtain:
\begin{equation}
r_\mathcal{D}(\theta,\theta',s) = \lambda v_0f_D(\theta')\max\big\{0,(e^{j\theta}-e^{j\theta'})\cdot \mathbf{t}(s;\theta,\theta')\big\} |\mathbf{b}'(s;\theta,\theta')|.
\label{rate_d}
\end{equation}

Having computed expressions for $r_{\mathcal A}$, $r_{\mathcal B}$, $r_{\mathcal C}$, and $r_{\mathcal D}$, we finally define one last family of transition rates that will be used in subsequent derivations. First, given the value of $\Theta_i=\theta$ as before, we define the family of events
\begin{equation*}
\hat{\mathcal{D}}(\theta)=\{\hat{\mathcal{D}}(\theta,\theta',\mathbf{r}'),~ \theta' \in [-\pi,\pi),~\mathbf{r}' \in \mathcal{F}\},
\end{equation*}
where $\hat{\mathcal{D}}(\theta,\theta',\mathbf{r}')$ is the event 
that the buffering ends because, at time $T_i$, an eligible node appears
at position $\mathbf{r}'$ on the curve $\textbf{b}(s;\theta,\theta')$, 
traveling in direction $\theta'$. Also we define the transition
rates
\begin{equation*}
r_{\hat{\mathcal{D}}}(\theta,\theta',\mathbf{r}')d\theta' dA' dt= P(\hat{\mathcal{D}}(\theta,\theta',\mathbf{r}'),\Delta_i  =t|\Theta_i=\theta), 
\quad \theta,~\theta' \in [-\pi,\pi),~\mathbf{r}' \in \mathcal{F}.
\end{equation*}
Observe that $r_{\hat{\mathcal{D}}}(\theta,\theta',\mathbf{r}')$ is
simply a different representation  of
$r_\mathcal{D}(\theta,\theta',s)$,
and it can be easily recovered from knowing
$r_\mathcal{D}(\theta,\theta',s)$
and $\textbf{b}(s;\theta,\theta')$.
Indeed, fixing $\theta$ and $\theta'$, the rate
$r_\mathcal{D}(\theta,\theta',s)$ 
only specifies the transition rates of eligible node arrivals but not 
their locations; these are provided by the 
function $\textbf{b}(s;\theta,\theta')$; on the other hand, 
the rate $r_{\hat{\mathcal{D}}}(\theta,\theta',\mathbf{r}')$ 
already contains this information.
In particular, we have
\begin{equation}
\int_{\hat{F}} r_{\hat{\mathcal{D}}}(\theta,\theta',\mathbf{r}')\, dA'
=\int_F r_\mathcal{D}(\theta,\theta',s)\, ds 
\label{eq45jk}
\end{equation}
for any $\hat{F} \subseteq \mathcal{F}$, with $F=\{s:b(\theta,\theta',s) \in \hat{F}\}\subseteq [0,1]$. 

In our numerical calculations later or, we calculate the rate 
$r_{\hat{\mathcal{D}}}(\theta,\theta',\mathbf{r}')$, for a specific pair $\theta,\theta'$ 
and for all $\textbf{r}'$, as follows. First, we discretize $s$, defining $N$ values $s_i$, $i=1,2,\dots,N$ to cover $[0,1]$, each associated with an interval of length $\delta s_i$, $i=1,2,\dots,N$, the intervals partitioning $[0,1]$. Then, we discretize $\mathbf{r}'$, defining $M$ values $\mathbf{r}'_j$, $j=1,2,\dots,M$, each associated with an area $\delta A_j$, the areas partitioning $\mathcal{F}$. We map each $s_i$ to the location $\mathbf{r}'_j$ nearest to $\mathbf{b}(\theta,\theta',s_i)$, and we denote the resulting
map by $\mathbf{r}'_m(\theta,\theta',\cdot)$. And setting
\begin{equation*}
r_{\hat{\mathcal{D}}}(\theta,\theta',\textbf{r}_j)=\frac{1}{\delta A_j} \sum_{s_i:\textbf{r}_j=\mathbf{r}'_m(\theta,\theta',s_i)} r_\mathcal{D}(\theta,\theta',s_i)\delta s_i,
\end{equation*}
we have a discretized version of (\ref{eq45jk}).

\subsection{Aggregate rates}

We also define the \textbf{aggregate rates} $r_\mathcal{A}(\theta)$,  $r_\mathcal{B}(\theta)$, $r_\mathcal{C}(\theta)$, $r_\mathcal{D}(\theta)$, and $r(\theta)$ as follows:
\begin{eqnarray}
r_\mathcal{A}(\theta)&=&\int_{-\pi}^\pi r_\mathcal{A}(\theta,\theta')\, d\theta',\nonumber \\
r_\mathcal{B}(\theta)&=&\int_{-\pi}^\pi \int_{\mathcal{F}} r_\mathcal{B}(\theta,\theta',\mathbf{r}')\,dA' d\theta', \nonumber \\
r_\mathcal{C}(\theta)&=&\int_{-\pi}^\pi  \int_{\mathcal{F}} r_\mathcal{C}(\theta,\theta',\mathbf{r}')\,  dA' d\theta',\label{eq34gh} \\
r_\mathcal{D}(\theta)&=&\int_{-\pi}^\pi  \int_0^1 r_\mathcal{D}(\theta,\theta',s)\,ds d\theta', \label{eq34gh2}\\
r(\theta)&=&r_\mathcal{A}(\theta)+r_\mathcal{B}(\theta)+r_\mathcal{C}(\theta)+r_\mathcal{D}(\theta). \label{eq4ho9}
\end{eqnarray}
The interpretation of the first four rates above is that, each one of them,
multiplied by $dt$, is the infinitesimal (conditional) probability that an 
event from the corresponding family will occur 
after a time $\Delta_i\in[0,dt]$, 
given that the packet is traveling in direction $\Theta_i=\theta$. 
And the last one rate, $r(\theta)$, multiplied by $dt$, gives 
the probability that the buffering of stage $i$ will end after
a time $\Delta_i\in [0,dt]$, given that $\Theta_i=\theta$.

Observe that we must have 
\begin{equation*}
r_0=r_\mathcal{A}(\theta)+r_\mathcal{B}(\theta),
\end{equation*}
as the union of the events belonging to the families $\mathcal{A}$ and $\mathcal{B}$ is the event that node $A$ changes direction, which happens with rate $r_0$. Therefore,
\begin{equation}
r(\theta)=r_0+r_\mathcal{C}(\theta)+r_\mathcal{D}(\theta).
\label{eq45uu}
\end{equation}

\subsection{Time-Invariance Approximation and consequences}
\label{subsection77jj}

The transition rates of the events in the four families defined in 
Section~\ref{section6hg} are not independent of the duration $\Delta_i$ 
of the buffering stage. Intuitively, as time progresses, memory accumulates, 
and the probability of each of them occurring changes. 
As this fact significantly complicates the analysis required for computing the probability that a specific one of these events occurs, we adopt the following simplifying assumption:

\medskip

\noindent
\textbf{Time-Invariance Approximation:} \emph{For each incremental event 
$E \in \mathcal{U}(\theta)\cup \hat{\mathcal{D}}(\theta)$, and any time
$t\geq 0$, as $\delta t\downarrow 0$ we have:
\begin{equation*}
\frac{1}{\delta t}
P(E, t \leq \Delta_i \leq t+\delta t|\Delta_i \geq t, \Theta_i=\theta)
=\frac{1}{\delta t}
P(E,~0 \leq \Delta_i \leq \delta t| \Theta_i=\theta) + o(1).
\end{equation*}}

\smallskip

Intuitively, under this approximation, the probability that the buffering will 
end in a specific manner does not change as the stage progresses, but it is 
equal to the probability that this will happen right at the moment when the 
buffering starts (and the mobility process has been restarted,
due to the Second-Order Approximation). 
In particular, integrating the above expression over all
$E \in \mathcal{U}(\theta)\cup \hat{\mathcal{D}}(\theta)$ implies
that $\Delta_i$ is memoryless, in that
$$P(t \leq \Delta_i \leq t+dt|\Delta_i \geq t,\Theta_i=\theta)
=P(0 \leq \Delta_i \leq dt|\Theta_i=\theta)=r(\theta)dt,$$
therefore, under the Time-Invariance Approximation, 
each $\Delta_i$ is exponentially distributed
\cite{ross1} 
with rate $r(\theta)$.

Furthermore, the Time-Invariance Approximation makes it possible
to obtain ``time-averaged'' versions of the expressions
for the rates in~(\ref{eq:RA})--(\ref{eq:RD}).
For example, 
adopting the same slight abuse of notation as before,
for any event $\mathcal{A}(\theta,\theta')$ we have
\begin{eqnarray*}
P(\mathcal{A}(\theta,\theta'),	\Delta_i=t|\Theta_i=\theta)
&=& P(\mathcal{A}(\theta,\theta'), \Delta_i=t, 	\Delta_i \geq t|\Theta_i=\theta) \\
&=& P(\mathcal{A}(\theta,\theta'), \Delta_i=t|\Delta_i \geq t,\Theta_i=\theta) 	P(\Delta_i \geq t|\Theta_i=\theta) \\
&=& P(\mathcal{A}(\theta,\theta'),\Delta_i=0	|\Theta_i=\theta)P(\Delta_i \geq t|\Theta_i=\theta)\\
&=& r_\mathcal{A}(\theta,\theta')d\theta'dt \exp\{-t r(\theta)\},
\end{eqnarray*}
where the third equality follows from the Time-Invariance Approximation,
and the last equality from the definition of $r_\mathcal{A}$ and the
fact that, conditional of $\Theta_i=\theta$,
$\Delta_i$ is exponential with rate $r(\theta)$. 
Integrating over $0\leq t<\infty$,
we then obtain
$$P(\mathcal{A}(\theta,\theta')|\Theta_i=\theta)
= r_\mathcal{A}(\theta,\theta') d\theta' \int_0^\infty 
	\exp\{-t r(\theta)\} dt\\ 
= \frac{r_\mathcal{A}(\theta,\theta') d\theta'}{r(\theta)}.
$$
Working in the same manner for the other families, we can arrive at similar results. Summarizing,  
\begin{eqnarray}
P(\mathcal{A}(\theta,\theta')|\Theta_i=\theta)&=&\frac{r_\mathcal{A}(\theta,\theta') d\theta'}{r(\theta)}, \label{eq34w1}\\
P(\mathcal{B}(\theta,\theta',\mathbf{r}')|\Theta_i=\theta)&=&\frac{r_\mathcal{B}(\theta,\theta',\mathbf{r}')d\theta' dA'}{r(\theta)}, \label{eq34w2}\\
P(\mathcal{C}(\theta,\theta',\mathbf{r}')|\Theta_i=\theta)&=&\frac{r_\mathcal{C}(\theta,\theta',\mathbf{r}')d\theta'dA'}{r(\theta)}, \label{eq34w3}\\
P(\mathcal{D}(\theta,\theta',s)|\Theta_i=\theta)&=&\frac{r_\mathcal{D}(\theta,\theta',s)d\theta'ds}{r(\theta)},\label{eq34w4}\\
P(\hat{\mathcal{D}}(\theta,\theta',s)|\Theta_i=\theta)&=&
\frac{r_{\hat{\mathcal{D}}}(\theta,\theta',\mathbf{r}')d\theta'dA'}{r(\theta)}.\label{eq34w5}
\end{eqnarray}

\newpage

\section{Performance Metrics}
\label{section1t7}

In this section we will derive expressions for the long-term average packet speed and cost induced by our RR on this network model. These will be expressed in terms of the invariant distribution of an appropriately defined Markov chain. The following is the last technical assumption we need to impose on the potential function:

\medskip

\noindent
\textbf{Assumption 4:} The value of 
$U(-\pi, \textbf{r})$ is equal to a constant 
$K$ for all $\mathbf{r} \in \mathcal{F}$. 

\medskip

Coupled with the monotonicity of $U(\theta,\mathbf{r})$,
this assumption simply states that the direction $\theta=-\pi$ 
is uniformly the worst, irrespective of the location $\mathbf{r}$ 
of a candidate neighbor. Note, however, that the 
behavior of $U(\theta,\mathbf{r})$ as a function of $\theta$ can strongly 
depend on $\mathbf{r}$, so that `good' locations can be favored, in terms 
of the potential assigned to them, as long as nodes at those locations are 
not traveling in direction $-\pi$. Therefore, this assumption is clearly
not significantly restrictive. In technical terms, it will be used to establish
the irreducibility of the chain $\{S_i\}$ defined below. As should become
evident from the analysis, this assumption could be relaxed, but at the cost
of significantly complicating some of the arguments involved, so we will
not pursue this direction further.

\subsection{The Markov chain}

We define the {\bf state} $S_i$ associated with each stage $i\geq 1$, by $S_i  \triangleq (\Theta_i,(X_{W,i},Y_{W,i}))$ if $i$ is a wireless transmission stage, and by $S_i  \triangleq (\Theta_i,(0,0))$ if $i$
is a buffering stage. The associated {\bf state space} in which each $S_i$ takes values is $\mathcal{S}=\mathcal{S}_W \cup \mathcal{S}_B$, where the \textbf{transmission state space}
\begin{equation*}
\mathcal{S}_W \triangleq (-\pi,\pi) \times (\mathcal{F}-\{\mathbf{0}\})
\end{equation*}
and the \textbf{buffering state space}
\begin{equation*}
\mathcal{S}_B \triangleq [-\pi,\pi) \times \{\mathbf{0}\}.
\end{equation*} 
By Assumptions 1 and 4, a node $A$ traveling in direction $-\pi$ will never receive a packet from a node $B$, irrespective of its location $\mathbf{r}$ and node $B$'s traveling direction, therefore the pairs $(-\pi,\mathbf{r})$ with $\mathbf{r}\neq\mathbf{0}$ are not included in $\mathcal{S}_W$. 

Observe that, due to the Second-Order Approximation, the process $\{S_i,~i=1,2,\dots\}$ forms a Markov chain: If $S_i=(\theta,\mathbf{0})$, i.e., $i$ is a buffering stage, then at the start of that stage the complete mobility model was restarted, except that the carrier $A$ kept its travel direction $\theta$ and its FR did not contain nodes with a potential higher than that of $A$, i.e., $U(\theta,\textbf{0})$. Likewise, if $S_i=(\theta_i,\mathbf{r})$ with $\mathbf{r} \neq \mathbf{0}$, i.e., in stage $i$ the packet is transmitted from a node $B$ to a node $A$ located at $\mathbf{r} \in \mathcal{F}(B)$, then, at the moment $A$ received the packet, the whole mobility model was again restarted, except that $A$ kept its travel direction $\theta$ and all nodes with potential higher than $U(\theta,\textbf{r})$ were expunged from $\mathcal{F}(A)\cap \mathcal{F}(B)$. In both cases, the complete information remaining about the network is captured in the current state. 

The distribution of the chain $\{S_i\}$ may be described as follows. We assume that $S_1=s\in{\cal S}$ is an arbitrary initial state, and for each $i$, given $S_i=(\theta,\mathbf{r})$, the chain moves to a state $S_i=(\theta',\mathbf{r}')$ according to the following family of conditional distributions, as derived in the previous section: If $\mathbf{r}=\mathbf{r}'=\mathbf{0}$, the conditional density of $S_{i+1}$ is
$$K_{BB}(\theta;\theta') = \frac{r_\mathcal{A}(\theta,\theta')}{r(\theta)};$$
if $\mathbf{r}=\mathbf{0}$ and $\mathbf{r}'\neq\mathbf{0}$, the conditional density of $S_{i+1}$ is
$$K_{BW}(\theta;\theta',\mathbf{r}') = \frac{r_\mathcal{B}(\theta,\theta',\textbf{r}') 
+ r_\mathcal{C}(\theta,\theta',\textbf{r}') + r_{\hat{\mathcal{D}}}(\theta,\theta',\textbf{r}')}{r(\theta)};$$
if both $\mathbf{r}$ and $\mathbf{r}'$ are nonzero, then the conditional density of $S_{i+1}$ is 
$$K_{WW}(\theta,\textbf{r}; \theta', \textbf{r}' ) =g(\theta',\textbf{r}';\theta,\textbf{r}),$$
where the function $g$ is given in~(\ref{eq451d7}); and finally, if $\mathbf{r}\neq\mathbf{0}$ and $\mathbf{r}'=\mathbf{0}$, then the conditional density of $S_{i+1}$ is 
$$K_{WB} (\theta,\textbf{r}; \theta')= \delta(\theta'-\theta) P_E(\theta,\textbf{r}).$$
In the sequel we refer to $K_{BB}(\theta;\theta')$, $K_{BW}(\theta; \theta', \textbf{r}')$, $K_{WW}(\theta,\textbf{r}; \theta', \textbf{r}' )$, and $K_{WB} (\theta,\textbf{r}; \theta')$ as the \textbf{kernel functions}, since they can be used to fully specify the transition kernel of the chain $\{S_i\}$.

\subsection{Ergodicity}

In this section we establish that, 
under the Second-Order Approximation,
the Time-Invariance Approximation, and Assumptions~1--4, the Markov chain
is ergodic, with a unique invariant distribution $\pi$, to which it 
converges at a geometric rate.

Let ${\cal L}_1$ denote the Lebesgue measure on $[-\pi,\pi)$,
${\cal L}_2$ denote the Lebesgue measure on ${\cal F}$, and
$\delta_\mathbf{0}$ be the point mass at point $\mathbf{0}=(0,0)\in\R^2$.
We write $\psi$ for the measure 
$\psi={\cal L}_1\times\delta_\mathbf{0}+{\cal L}_1\times{\cal L}_2$, 
defined on the state space ${\cal S}$,
equipped with the usual Borel $\sigma$-field. Our first result
describes the long-term behavior of the chain $\{S_i\}$,
and its consequences are stated in detail after that;
see \cite{meyn-tweedie:book2} for some relevant background
on Markov chains. Theorem~\ref{t:ergodicity} is proved
in the Appendix.

\medskip

\begin{Theorem}
\label{t:ergodicity}
Under the Second-Order Approximation,
the Time-Invariance Approximation, and Assumptions~1--4, 
the Markov chain is $\psi$-irreducible, aperiodic, 
and uniformly ergodic on the state space ${\cal S}$,
with a unique invariant measure $\pi$ to which it
converges uniformly geometrically fast. In particular:
\begin{enumerate}
\item
There are constants $B<\infty$ and $\rho\in(0,1)$ such that,
for any initial state $s\in{\cal S}$,
$$|P(S_n\in A|S_1=s)-\pi(A)|\leq B\rho^n,$$
for all $n\geq 1$ and any (measurable) set $A\subset{\cal S}$.
\item
For any (measurable) function $F:{\cal S}\to\R$
with $E_\pi[|F(S)|]<\infty$,
as $n\to\infty$ we have, with probability one,
$$\frac{1}{n}\sum_{i=1}^n F(S_i)\to E_\pi[F(S)],$$
for any initial state $s\in{\cal S}$, where $S\sim\pi$.
\end{enumerate}
\end{Theorem}
   
\medskip

An important ingredient in the proof of Theorem~\ref{t:ergodicity}
is the following domination condition, which 
will be verified in the Appendix.
Intuitively, Lemma~\ref{lem:doeblin}
says that, irrespective of the current state,
with probability at least $\epsilon$ the chain will be in a uniformly
distributed buffering state after two time steps.

\medskip

\begin{Lemma} \emph{(Doeblin condition)}
\label{lem:doeblin}
Let $\mu$ denote the measure ${\cal L}_1\times\delta_{\mathbf{0}}$ 
on ${\cal S}$. There is an $\epsilon>0$ such that
for any (measurable)
$A \subset \mathcal{S}$ and any $s \in \mathcal{S}$, we have:
\begin{equation*}
P(S_{i+2} \in A | S_i=s) 
\geq \epsilon \mu (A).
\end{equation*}
\end{Lemma}

\medskip

Another ingredient of the proof of the $\psi$-irreducibility
part of Theorem~\ref{t:ergodicity} is provided by the following
one-step reachability bound. Lemma~\ref{lem:irred} is proved in 
the Appendix.

\medskip

\begin{Lemma} 
\label{lem:irred}
Let $\mu'$ denote the measure ${\cal L}_1\times{\cal L}_2$ on
${\cal S}_W$. For any (measurable) $A\subset{\cal S}_W$
with $\mu'(A)>0$ there are 
$-\pi\leq \theta'_1<\theta'_2<\pi$ such that,
\begin{equation}
P(S_{i+1} \in A | S_i = (\theta,\mathbf{0}))>0, \quad \mbox{for all}
\; \theta \in (\theta'_1,\theta'_2).
\label{eq78j}
\end{equation}
\end{Lemma}

\medskip

The main implications of Theorem~\ref{t:ergodicity}
for our results are stated in the following corollary,
which is proved in the Appendix. In order to state
it we need some additional definitions. 
Given an arbitrary
state $S_1=s=(\theta,(x_{W},y_{W}))$ in ${\cal S}$,
let $\Delta_1$ be exponentially distributed with rate $r(\theta)$
if $(x_W,y_W)=\mathbf{0}$, and $\Delta_1=0$ otherwise.
Similarly, for each $i\geq 2$, given 
$(S_1,\ldots,S_{i-1},S_i=(\theta,(x_W,y_W)))$ 
and $(\Delta_1,\ldots,\Delta_{i-1})$,
let $\Delta_i$ have the same distribution as $\Delta_1$
given $(\theta,(x_W,y_W))$. 
Then $\{\bar{S}_i=(\Theta_i,(X_{W,i},Y_{W,i}),\Delta_i)\}$
defines a new Markov chain, on the state space:
$$\bar{\cal S}=
\big([-\pi,\pi) \times \{\mathbf{0}\}\times[0,\infty)\big)
\cup
\big((-\pi,\pi) \times (\mathcal{F}-\{\mathbf{0}\})\times\{0\}\big).
$$
Now suppose
$S=(\Theta,(X_{W},Y_{W}))$ has distribution $\pi$ 
and let $\Delta$ be defined as before, conditional
on $S$. Write $\bar{\pi}$ for the induced joint
distribution of 
$\bar{S}=(\Theta,(X_{W},Y_{W}),\Delta)$ on $\bar{\cal S}$.

\medskip

\begin{Corollary}
\label{cor:ets}
For any initial state $S_1=s$, $\Delta_1=\delta$, 
the following ergodic theorems hold with probability one:
\begin{eqnarray*}
\limn \frac{1}{n}\sum_{i=1}^n X_{W,i}&=& E_{\pi}(X_{W}), \\
\limn \frac{1}{n}\sum_{i=1}^n C_{i}&=& 
	E_{\pi}(C)
	\;=\;
	E_{\pi}(C(X_W,Y_W)),\\
\limn \frac{1}{n}\sum_{i=1}^n \Delta_{i}&=& E_{\bar{\pi}}(\Delta), \\
\limn \frac{1}{n}\sum_{i=1}^n X_{B,i}&=& E_{\bar{\pi}}(X_B)
	\;=\;v_0 E_{\bar{\pi}}
	(\Delta\cos\Theta),
\end{eqnarray*}
where $(\Theta,(X_W,Y_W),\Delta)\sim\bar{\pi}$ so that,
in particular,
$(\Theta,(X_W,Y_W))\sim\pi$.
\end{Corollary}

\medskip

As the final step of our analysis, we provide expressions for 
the performance metrics defined in Section~\ref{subsection1}. 
The following results, stated without proof, are immediate 
consequences of Corollary~\ref{cor:ets}.

\medskip

\begin{Corollary}
For any initial state $S_1=s$, $\Delta_1=\delta$, the limits defining
the performance metrics $V_p$ and $C_p$ in~(\ref{eq1})
and~(\ref{eq2}), respectively, exist with probability one,
and are given by:
\begin{eqnarray}
V_p &=&
	\frac{E_{\bar{\pi}}(X_{W}+v_0\Delta\cos\Theta)}
	{E_{\bar{\pi}}(\Delta)},
	\label{eq3}\\
C_p &=&
	\frac{E_{\bar{\pi}}(C(X_W,Y_W))}
	{E_{\bar{\pi}}(X_{W}+v_0\Delta\cos\Theta)},
	\label{eq5}
\end{eqnarray}
where $(\Theta,(X_W,Y_W),\Delta)\sim\bar{\pi}$.
\end{Corollary}

\medskip

Some details regarding the numerical computation of expectations
under $\pi$ and $\bar{\pi}$ are given in Section~\ref{s:pi} of the 
Appendix.

\newpage

\section{Numerical Results}
\label{section_results}

In this section, we compare the approximate results for the performance
metrics $V_p$ and $C_p$ obtained in previous sections, with corresponding
simulation results for specific choices of the FR ${\cal F}$ and the 
potential function $U(\cdot,\cdot)$.

\subsection{Setting}
\label{section67hj}
\begin{figure}[ht!]
\centering
\includegraphics{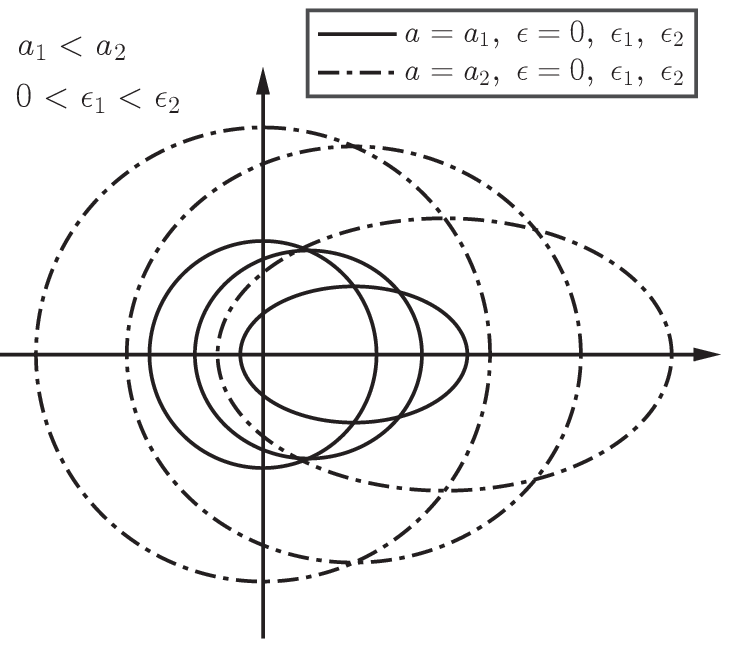} \hspace{0.3in}
\includegraphics{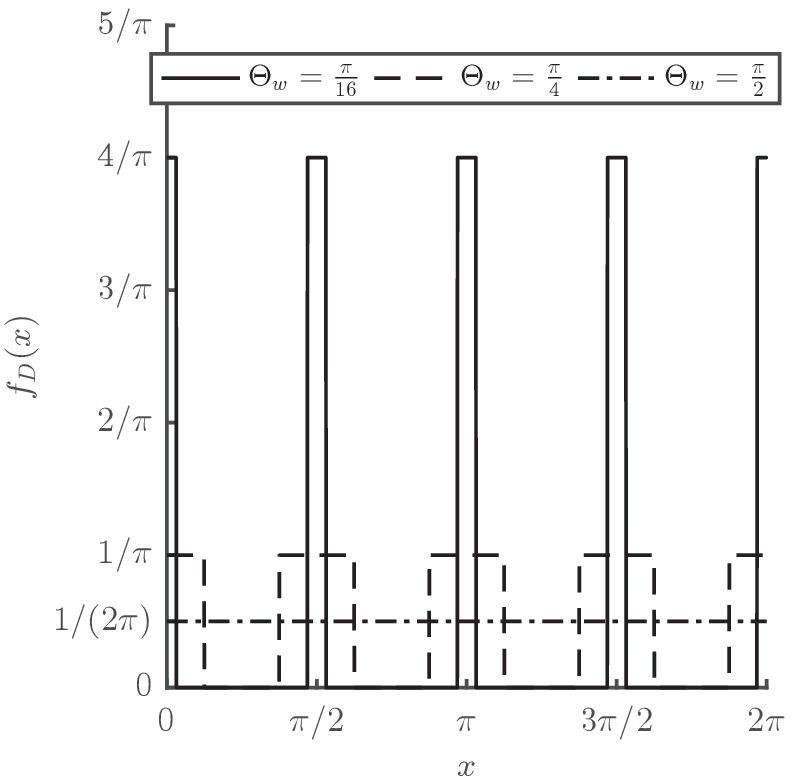}
\caption{Left: The boundary function $b(\phi)$ for two different
values $a_1<a_2$ of the parameter $a$, and three different
values $0<\epsilon_1<\epsilon_2$ of the parameter $\epsilon$.
Right:
The density $f_D(\cdot)$ for three different values of the parameter $\Theta_w$.}
\label{fig111as}
\end{figure}

We consider the potential function
\begin{equation*}
U(\theta,\mathbf{r})=-|\theta|,\;\;\;\theta\in[-\pi,\pi),
\end{equation*}
so the packet constantly tries to find nodes with a 
good travel direction, regardless of their relative location, provided, 
of course, they are within the FR. The FR ${\cal F}$ we consider
is specified by the boundary function
\begin{equation*}
b(\phi)=\frac{a(1-\epsilon^2)}{1-\epsilon \cos \phi},
\end{equation*}
for some $a>0$, $\epsilon  \in (0,1)$, so that
${\cal F}$ is an ellipse whose major axis, of length $2a$,
is along the $x$-axis, its left focus is 
at the origin, and its eccentricity is $\epsilon$. 
The boundary function is drawn in Fig.~\ref{fig111as} 
for a few different choices of the parameters $a$ and $\epsilon$. 
Note that larger values of $a$ make the routing protocol more aggressive 
in finding nodes to send the packet to, whereas larger values of
$\epsilon$ make the routing protocol more selective regarding the 
relative locations of neighboring nodes. In the important 
special case $\epsilon=0$, ${\cal F}$ is a circle of radius $a$, 
so that the routing protocol gives the packet to any node with 
a direction better than that of the current holder, as long as the 
two nodes are within distance $a$ of each other.
This choice of boundary function may be the only possible 
if the nodes only know their travel directions but not their
relative locations, 
and they can exchange packets whenever they are within 
a communication radius $a$ of each other.  

Finally, we assume that the transmission cost is quadratic, 
$C(\mathbf{r})=|\mathbf{r}|^2$, and for the density of the
travel direction $f_D(\cdot)$ we take
\begin{equation*}
f_D(x)=\begin{cases} \frac{1}{4\Theta_w}, & \left|x-k\frac{\pi}{2}\right| < \frac{\Theta_w}{2}~~\mbox{for some } k \in \Z, \\ 0, & \mbox{elsewhere,} \end{cases}
\end{equation*}
where $\Theta_w \in [0,\pi/2]$. Therefore, $f_D(\cdot)$ is positive and constant 
on four intervals centered in the directions of the positive and 
negative $x$- and $y$-axes, whereas outside of these ranges $f_D(\cdot)$ is zero. 
For $\Theta_w=\pi/2$, the density is the uniform density 
$f_D(x)=\frac{1}{2\pi}$, for all directions $x$. 
At the other extreme, small values of $\Theta_w$ model situations 
in which all nodes move along the direction of one of two perpendicular axes; this would happen, for example, with a vehicular network of nodes moving 
in a rectangular road grid. In Fig.~\ref{fig111as} we also plot the 
density $f_D(\cdot)$ for three different values of the parameter $\Theta_w$.
In Table~\ref{table45gg} we collected all the quantities used in the 
calculations of this section, along with their default values; these 
values are used in all computations, unless explicitly stated otherwise.

\begin{table}[ht!]
\begin{center}
\caption{Quantities and their default values used in Section~\ref{section_results}}
\label{table45gg}
\begin{tabular}{|c|c|c|} \hline
Quantity & Symbol & Default value \\ \hline 
Node density & $\lambda$ & $1$\\ 
Direction density & $f_D(x)=\begin{cases} \frac{1}{4\Theta_w}, & \left|x-k\frac{\pi}{2}\right| < \frac{\Theta_w}{2}~~\mbox{for some } k \in \Z, \\ 0, & \mbox{elsewhere} \end{cases}$ & $\Theta_w=\frac{\pi}{2}$ (uniform)\\ 
Node speed & $v_0$ &  $1$ \\
Node turning rate & $r_0$ & $1$ \\ 
Transmission cost & $C(\mathbf{r})=|\mathbf{r}|^2$ & N/A \\
Boundary function & $b(\phi)=\frac{a\left(1-\epsilon^2\right)}{1-\epsilon \cos \phi}$, $~\phi \in [-\pi,\pi)$& $a=1$, $\epsilon=0.7$ \\ 
Potential & $U(\theta,\mathbf{r})=-|\theta|$, $~\mathbf{r} \in \mathcal{F}(A)$,~$\theta \in [-\pi,\pi)$ & N/A \\ \hline
\end{tabular}
\end{center}
\end{table}

\subsection{Results}
\label{s:results}

\begin{figure}[ht!]
\centering
\includegraphics{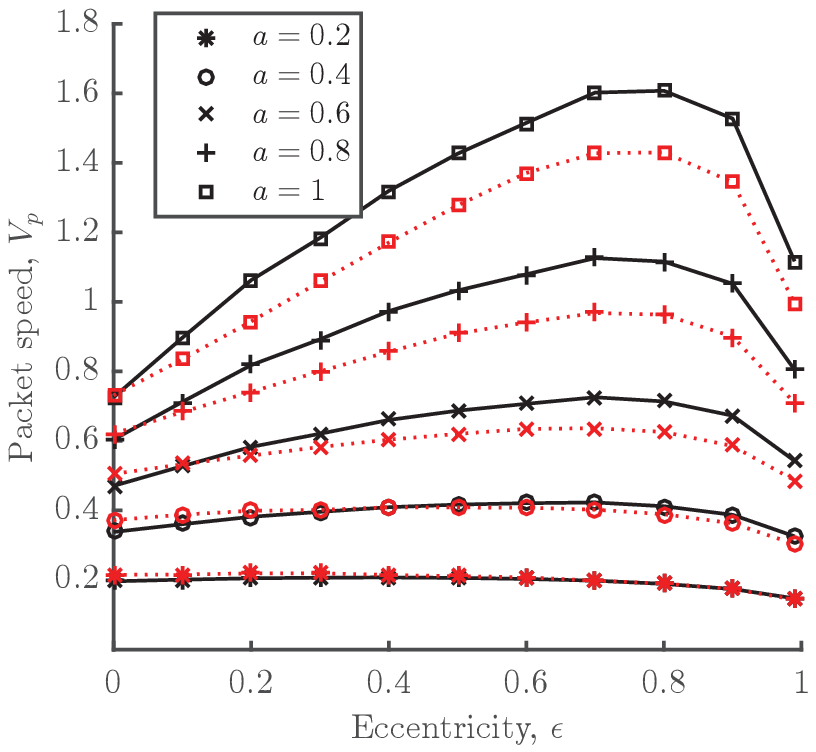} \hfill
\includegraphics{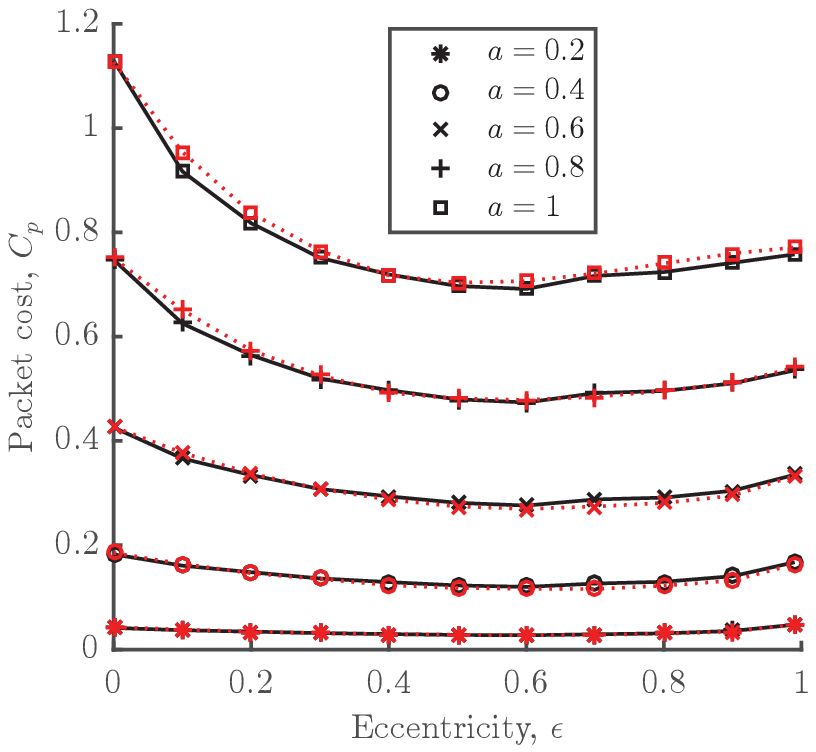} 
\caption{$V_p$ and $C_p$ versus $a$ and $\epsilon$.
Solid black lines depict our analytical results, 
and dotted red lines depict simulation results.}
\label{experiment1}
\end{figure}

Fig.~\ref{experiment1} shows the effects of the shape of the FR 
(as the eccentricity $\epsilon$ and the half-axis length $a$ vary)
on the packet speed $V_p$ and the packet cost $C_p$.
Here, and in all subsequent figures, the results obtained from our 
earlier analysis are shown as solid black lines, and the corresponding 
simulation results are shown as dotted red lines.

Observe that, as the half-axis length $a$ increases, the packet speed increases but so does the packet cost; this exemplifies the fundamental trade-off between these two metrics. The increase in the speed as $a$ gets larger is because it becomes more likely for a node with a good travel direction to be available when the carrier changes direction to a bad one, and also (when $\epsilon>0$) because that node is farther ahead on the average; for the same reason, and also because the transmission cost function is quadratic, $C_p$ also increases as $a$ increases. In fact, as the figure suggests, we expect that when $\epsilon>0$, the speed diverges to infinity as $a$ increases, since the expected progress per wireless transmission increases  with $a$. On the other hand, the cost diverges to infinity, as $a$ increases, even when $\epsilon=0$. 

Regarding the effects of the eccentricity $\epsilon$, observe that, starting from 
$\epsilon=0$ and increasing it, initially leads to higher speed and lower cost.
This is natural, as the value $\epsilon=0$ corresponds to a circular FR, 
therefore neighboring nodes whose relative position is towards the 
positive $x$-axis are not given preference; this inefficiency is rectified 
as $\epsilon$ initially increases. However, increasing $\epsilon$ past 
$\epsilon\approx 0.6$ actually leads to an increase in the cost. Indeed, 
if the FR is \emph{too} elliptical, it often happens that the packet is 
transmitted to nodes that are too far away from the current carrier, 
albeit with an excellent relative position, although there were other 
nodes that were much closer to the carrier with a relative position 
almost as good; as the cost is quadratic, this inevitably increases the packet 
cost. Large values of the eccentricity also hurt the speed because, when 
$\epsilon$ increases, the area of the FR is reduced 
(the exact formula is $|\mathcal{F}(A)|=\pi a^2\sqrt{1-\epsilon^2}$),
and the packet spends more time traveling towards relatively bad 
directions on the buffers of nodes.

\begin{figure}[ht!]
\centering
\includegraphics{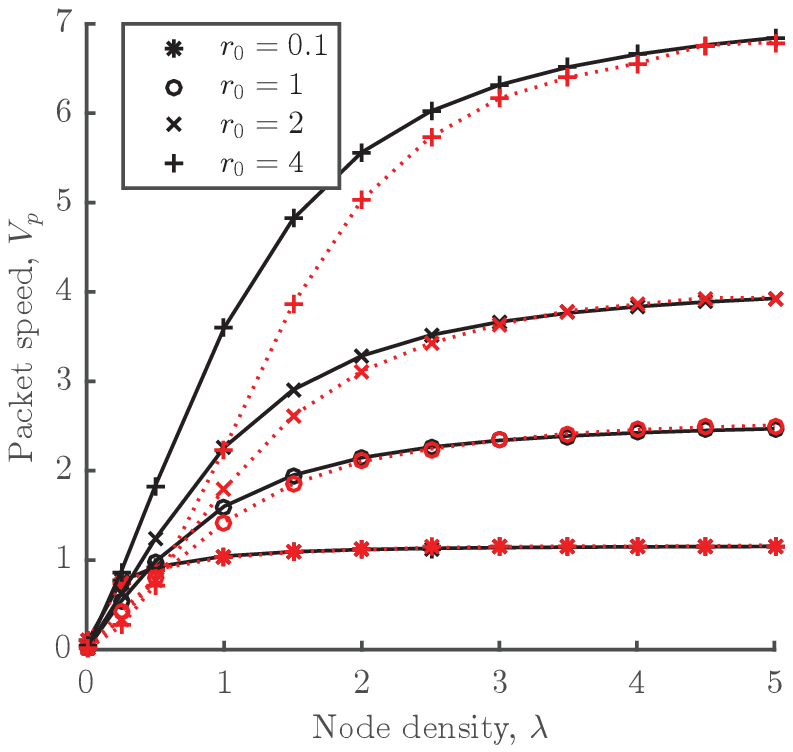} \hfill
\includegraphics{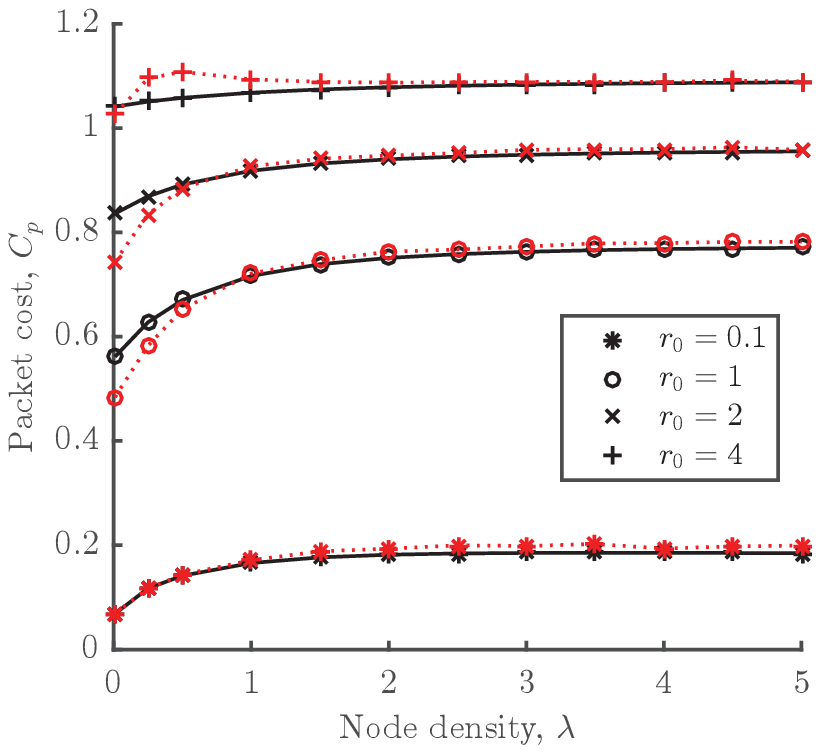} 
\caption{$V_p$ and $C_p$ versus $\lambda$ and $r_0$.
Solid black lines depict our analytical results, 
and dotted red lines depict simulation results.}
\label{experiment2}
\end{figure}

In Fig.~\ref{experiment2} we plot the values of $V_p$ and $C_p$ versus 
the two node parameters, namely, the node density $\lambda$ and the node 
turning rate $r_0$. 

Regarding the effects of $r_0$, we first observe that, when $r_0$ is very small, as long as the node density is not very small, the \emph{packet} speed is almost equal to the \emph{node} speed. Indeed, the packet stays with a node with a near-perfect direction for a significant amount of time, and in the infrequent cases when that node changes its direction, another one will be found within a relatively short time. Consequently, the packet cost is also very small. On the other hand, when $r_0$ is not very small, then, the larger $r_0$ is, i.e., the more frequently a node changes direction, the more frequent are the transmissions to nodes with better directions, and hence both the packet cost and the speed get larger; this effect on the speed crucially depends on the fact that transmissions are, on average, towards the direction of the positive $x$-axis, since, in this figure, we use the default value $\epsilon=0.7$.

Regarding the effects of $\lambda$, when $r_0$ is fixed and non-negligible, a low density $\lambda$ leads to low packet speed, as the packet spends extended periods of time traveling towards bad directions; on the other hand, a large node density means that the packet travels fast, due to frequent transmissions. However, this effect diminishes as, after a while, a node with near-perfect travel direction is guaranteed to exist within the FR whenever the current carrier changes its travel direction; therefore, increasing the density further has no effect. On the other hand, $C_p$ is near-constant as $\lambda$ changes. To understand this, compare the high-density regime with the low-density regime: In the first case, the travel of the packet consists of wireless transmissions and physical transports in the right direction. In the second case, it again consists of wireless transmissions and transports in the right direction, but also involves extended periods of transports in \emph{random} directions (which, on the average, produce no progress). The two cases differ significantly in their performance in terms of progress per unit time (i.e., the speed) but \emph{not} in terms of cost per unit distance (in the positive $x$-axis direction), as transports in random directions have an approximately zero net effect.

\begin{figure}[ht!]
\centering
\includegraphics{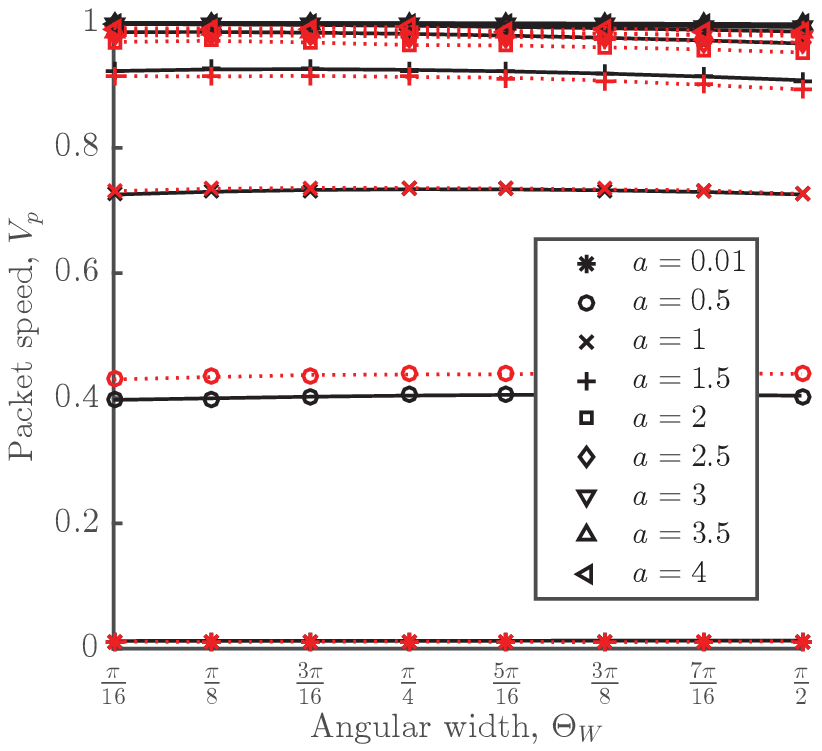} \hfill
\includegraphics{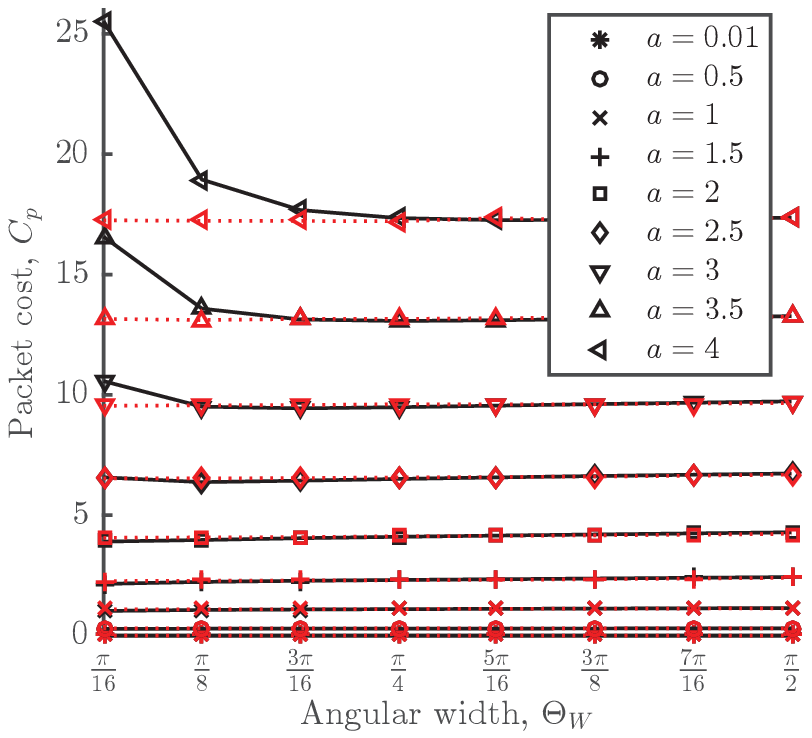} 
\caption{$V_p$ and $C_p$ versus $\Theta_w$ and $a$.
Solid black lines depict our analytical results, 
and dotted red lines depict simulation results.}
\label{experiment3}
\end{figure}

Finally, in Fig.~\ref{experiment3} we plot $V_p$ and $C_p$ versus the angular width $\Theta_w$ and the size $a$ of the half-length, when $\epsilon=0$, i.e., the FR ${\cal F}$ is a circular disk. 

Once again we observe that, as in Fig.~\ref{experiment1}, increasing $a$ increases both the speed and the cost. Indeed, the more nodes there are in the FR, the higher is the probability that, once the packet changes its travel direction, another node with a good travel direction will be available. However, in contrast to Fig.~\ref{experiment1}, as $a$ increases the effects on the speed tend to diminish; indeed, after some value of $a$, the probability that an eligible node with a good direction exists is invariably close to unity, and because now the eccentricity $\epsilon=0$, wireless transmissions have a zero net effect on the speed of the packet, which does not change with $a$.   

As for $\Theta_w$, its effects are much less pronounced: Both $V_p$ and $C_p$ change little with $\Theta_w$. This can be justified by observing that changing $\Theta_w$ does not make the directions with which the nodes travel overall better, only differently distributed; still, the effects of $\Theta_w$ on the performance metrics are remarkably small.

Finally, we note that the discrepancy between the simulation 
results and our analytical results is generally small and 
almost always modest. One exception is the 
setting of Fig.~\ref{experiment3} in the cases of both small values of $\Theta_w$ and large values of $a$, where the discrepancy is significant. However, this discrepancy is {\em not} due to the inaccuracy of our two simplifying approximations but, rather, due to accumulating numerical errors. Specifically, in this regime, the errors due to the discretizations used are large, because there is a non-negligible probability that there will be two or more nodes in the FR with the exact same (discretized) travel direction and, hence, potential once a packet arrives at a  new node or its current holder changes direction; note that our analysis assumes that the probability of this event is zero.
Excepting this case, the discrepancy between simulations and analysis remains modest, although it does increase with $r_0$ (cf.\ Fig.~\ref{experiment2}). Indeed, as $r_0$ increases, the Second-Order Approximation is invoked more frequently, and the estimated rates with which the FR encounters eligible nodes deviate more from the actual ones.

\newpage


\section{Conclusions} 
\label{section_conclusions}

In this work, we first introduced a mobile wireless DTN model in which nodes move on the infinite plane, according to a random waypoint mobility model, and a packet must travel to a destination located at an infinite distance away according to a routing rule that is using both wireless transmissions and physical transports on the buffers of nodes. The routing rule is defined in terms of a forwarding region and a potential function; specifying these leads to different versions of the routing rule. This model is quite general, notably including cases where the transmission cost depends on the direction of the transmission, arbitrary distributions for the direction of node travel, and a large variety of routing rules. 

In this setting, we defined two performance metrics: the speed with which 
the packet travels to its destination, and the rate with which the transmission cost is accumulated. We computed these performance 
metrics by adopting two simplifying approximations.
These approximations ensure that a simpler discrete-time Markov chain 
embedded in the system description can be analyzed using general
tools from Markov chain theory.
The assumptions are intuitive, and furthermore are shown to introduce 
modest errors, on the order of no more than 10\%, in the examples
considered in our numerical evaluations.

The present results help quantify the important trade-off that exists in mobile wireless DTNs between the speed with which packets travel to their destinations and the rate with which the transmission cost is accumulated. Also, the methodology we have developed may be extended in a variety of directions, e.g., 
to include the case where the velocity magnitude is not constant 
and the duration of time a node spends with a given travel 
direction depends on its velocity vector. Alternatively,
the present development may also be used as a starting point for more 
accurate analytical approaches, e.g., maintaining more memory in each 
stage of the Markov chain. Related work on much simpler 
settings~\cite{cheliotis1} suggests that dispensing with 
approximations altogether might be a formidable task.

An interesting potential application of our work is towards studying the performance of \emph{non}-delay-tolerant geographic routing protocols. Consider, for example, a simple RR with a circular FR of radius $R$, and such that when $R\rightarrow \infty$, with probability going to $1$ there will be an eligible node whenever a packet arrives at a new node; therefore, as $R$ increases physical transports become less and less frequent. Taking the limit as $R\rightarrow \infty$  of the cost $C_p$ readily gives the performance of a \emph{non}-delay-tolerant geographic routing protocol. 

Regarding future work, the present setting naturally leads to the problem of finding the \emph{best} routing rules, e.g., those that achieve Pareto optimal combinations of delays and costs. Tackling this problem with tools from genetic algorithms or multi-armed bandit theory jointly with tools from stochastic geometry might be a fruitful strategy. Also, the assumption that all nodes travel with the same speed is not crucial and could be relaxed; for example, we could use a more general model under which each node travels, independently of all others, with a speed that is constant between the times the node changes direction, but the speed changes when the direction changes, and speeds associated with consecutive trajectory segments are independent random variables following some given distribution. Likewise, we could assume that the distribution of the duration of time a node spends with a given travel direction does depend on that direction. On the other hand, two other aspects of our model, namely, the independence of node trajectories and the changes of each node's directions according to a Poisson process, cannot be easily relaxed, as that would introduce new sources of memory, making it more difficult to develop an accurate and tractable Markov chain model for the packet trajectory.

\newpage

\appendix

In the first section of the Appendix we give the proofs of the theoretical results in Section~\ref{section1t7}. In the other sections we provide details on the numerical evaluation of the performance metrics and intermediate results.

\subsection{Proofs}
Below we establish Theorem~\ref{t:ergodicity}, Lemmas~\ref{lem:doeblin} and~\ref{lem:irred}, and Corollary~\ref{cor:ets}. In several parts of the proofs where we need to invoke technical but quite standard arguments, some of the details are omitted.

\medskip

\begin{Proof}{ of Lemma~\ref{lem:doeblin}} 
It is obvious that it suffices to establish the result of the lemma for events of the form $A=A_0\times\{\mathbf{0}\}$, for $A_0\subset[-\pi,\pi)$. And by the uniqueness of Carath\'{e}odory extension, since the collection of all finite unions of intervals forms an algebra that generates the Borel $\sigma$-algebra of ${\cal S}$, it further suffices for $A_0$ to only consider closed intervals, $A_0=[\theta_1,\theta_2]$; see, e.g., \cite{royden:book,williams:book} for details. So in the rest of the proof we restrict attention to events $A$ of the form $A=[\theta_1,\theta_2]\times \{\mathbf{0}\}$.

Also note that, from the expressions derived in Section~\ref{s:transition}, it is simple to obtain the following bounds on the transition rates $r_{\mathcal{A}}$, $r_{\mathcal{B}}$, $r_{\mathcal{C}}$, $r_{\mathcal{D}}$, and on $r(\theta)$:
\begin{eqnarray}
r_0\epsilon_D \exp\left\{-\lambda |\mathcal{F}| \right\} 
&\leq& r_\mathcal{A}(\theta,\theta') \leq r_0 f_D(\theta'), \label{eq:RA_UL} \\
r_\mathcal{B} (\theta,\theta',\mathbf{r}')&\leq& r_0 \lambda f_D(\theta'), \nonumber\\ 
r_\mathcal{C} (\theta,\theta',\mathbf{r}')&\leq& r_0 \lambda f_D(\theta'),\nonumber \\ 
r_\mathcal{D} (\theta,\theta',s) & \leq & 2M_b \lambda v_0 f_D(\theta'), \nonumber\\
r(\theta) &\leq& r_0+ r_0\lambda |\mathcal{F}|+2M_b\lambda v_0. \label{eq92lk} 
\end{eqnarray}
We note that $\epsilon_D$ is a lower bound on $f_D$ defined in Section~\ref{sec78}.

Now, if $s$ is of the form $s=(\theta,\mathbf{0})$ for some $\theta\in[-\pi,\pi)$, then for any $-\pi\leq \theta_1<\theta_2<\pi$, 
\begin{equation*}
P\left(S_{i+1} \in [\theta_1,\theta_2] \times \{\mathbf{0}\}|
 S_i=(\theta,\mathbf{0}) \right)=\int_{\theta_1}^{\theta_2} 
\frac{r_\mathcal{A}(\theta,\theta')}{r(\theta)}\,d\theta',
\end{equation*}
so that, using the lower bound in~(\ref{eq:RA_UL}) and the upper bound in~(\ref{eq92lk}),
we have that, for some fixed constant $\delta_1>0$:
\begin{equation}
P\left(S_{i+1} \in [\theta_1,\theta_2]\times \{\mathbf{0}\} 
| S_i=(\theta,\mathbf{0}) \right) \geq \delta_1(\theta_2-\theta_1).
\label{eq45hp}
\end{equation}
Then, using the Markov property and applying~(\ref{eq45hp}) twice,
\begin{eqnarray}
P\left(S_{i+2} \in [\theta_1,\theta_2]\times \{\mathbf{0}\} 
	| S_i=(\theta,\mathbf{0}) \right) 
&\geq&
	P\left(S_{i+1}\in{\cal S}_B | S_i=(\theta,\mathbf{0})\right)
	\delta_1(\theta_2-\theta_1)
	\nonumber\\
&\geq&
	2\pi \delta_1^2(\theta_2-\theta_1).
\label{eq45hpb}
\end{eqnarray}

Similarly, if $s$ is of the form $s=(\theta,\mathbf{r})$
for some $\theta\in[-\pi,\pi)$ and $\mathbf{r}\in{\cal F}$,
then by the Markov property,
\begin{eqnarray*}
P\left(S_{i+2} \in [\theta_1,\theta_2]\times \{\mathbf{0}\} | S_i=(\theta,\mathbf{r} )\right)& \geq & P\left(S_{i+1} =(\theta,\mathbf{0}),~ S_{i+2} \in [\theta_1,\theta_2]\times \{\mathbf{0}\} | S_i=(\theta,\mathbf{r} )\right) \\
&=& P_E(\theta,\mathbf{r}) \int_{\theta_1}^{\theta_2} \frac{r_\mathcal{A}(\theta,\theta')}{r(\theta)}\, d\theta' \\
&\geq & \delta_1 P_E(\theta,\mathbf{r}) (\theta_2-\theta_1)\\
&= & \delta_1 \exp\{-E(N;\theta,\mathbf{r})\} (\theta_2-\theta_1),
\end{eqnarray*}
and recalling the expression for $E(N;\theta,\mathbf{r})$ 
in~(\ref{eq38}) we clearly have
$E(N;\theta,\mathbf{r})\leq\lambda|{\cal F}|$,
so that,
\begin{equation}
P\left(S_{i+2} \in [\theta_1,\theta_2]\times \{\mathbf{0}\} 
| S_i=(\theta,\mathbf{r} )\right)
\geq\delta_1\exp\{-\lambda|{\cal F}|\}(\theta_2-\theta_1).
\label{eq46hp}
\end{equation}

Combining~(\ref{eq45hpb}) and~(\ref{eq46hp}) yields the required result, with $\epsilon=\min\{2\pi \delta_1^2, \delta_1\exp\{-\lambda|{\cal F}|\}\}$.
\end{Proof}

\medskip

\begin{Proof}{ of Lemma~\ref{lem:irred}}
Since $A$ has positive Lebesgue measure, we can find
a rectangle
of the form 
$I=[\theta_1,\theta_2] \times [x_1,x_2] 
\times [y_1,y_2]\subseteq \mathcal{S}_W$
with a nonempty interior, such that
$\mu'(A\cap I)>0$.
The idea of the main argument here is to show that
there is a range of angles $(\theta'_1,\theta'_2)$ such that,
when the current packet holder 
travels with a direction in $(\theta_1',\theta_2')$, 
there is a strictly nonzero probability that there
are ineligible nodes in $[x_1,x_2] \times [y_1,y_2]$ 
that can become eligible by changing their travel direction to a better one within the range $[\theta_1,\theta_2]$. 

Since $U(\cdot,\cdot)$ is continuous, the image $U(I)$ of $I$ is  
a closed interval $[a,b]$. And since $I$ 
has a nonempty interior, we must have $a<b$ by Assumption~1. 
Also, by Assumptions~1 and~4, and noting that $\theta_1>-\pi$ 
in order to have $I\subseteq \mathcal{S}_W$,  we 
must have $b>a> U(-\pi,\mathbf{0})$. 

Next, pick $c,d$ such that $U(-\pi,\mathbf{0})<c<d<\min \{U(0,\mathbf{0}),a\}$,
and let $\theta'_1$ and $\theta'_2$ be such that 
$U(\theta'_1,\mathbf{0})=c$ and $U(\theta'_2,\mathbf{0})=d$; 
such angles are guaranteed to exist by the intermediate value theorem.
Also, observe that $U(\cdot,\cdot)$ is continuous on the compact set 
$[-\pi,0] \times [x_1,x_2] \times [y_1,y_2]$, so it is uniformly 
continuous there, which implies that there is 
a $\theta_B>-\pi$ with $U(\theta,\mathbf{r})<c$ for all 
$\theta \in [-\pi,\theta_B]$ and all $\mathbf{r} \in [x_1,x_2] 
\times [y_1,y_2]$.

Now take $(\theta',\mathbf{r}') \in I$ and 
$\theta \in (\theta'_1,\theta'_2)$ arbitrary. We will bound $r_{\mathcal{C}}(\theta,\theta',\mathbf{r}')$, given by (\ref{rate_c}), from below.
First note that $U(\theta',\mathbf{r}')>d$ and $U(\theta,\mathbf{0})<d$, therefore $U(\theta',\mathbf{r}')>U(\theta,\mathbf{0})$. Also, we have $U(\theta'',\mathbf{r}')<c<  U(\theta,\mathbf{0})$ for all $\theta'' \in [-\pi,\theta_B]$. Therefore, 
\begin{equation}
r_\mathcal{C}(\theta,\theta',\textbf{r}')\geq \lambda r_0f_D(\theta') \int_{-\pi}^{\theta_B}f_D(\theta'')\, d\theta''
\geq  \lambda r_0 \epsilon_D^2 (\theta_B+\pi), 
\label{eq78ja}
\end{equation}
where the second inequality follows by the fact that $f_D(\cdot)$ is assumed bounded below by $\epsilon_D$ (cf. Section~\ref{sec78}). Also recall that $r(\theta)$ is bounded above as in~(\ref{eq92lk}).

We are now ready to prove the inequality (\ref{eq78j}). For any $\theta \in (\theta'_1,\theta'_2)$, where the interval $(\theta'_1,\theta'_2)$ is chosen above, 
\begin{equation}
P(S_{i+1} \in A | S_i = (\theta,\mathbf{0}))\geq  
P(S_{i+1} \in A\cap I | S_i = (\theta,\mathbf{0}))\geq  
\int_{A\cap I} 
\frac{r_\mathcal{C}(\theta,\theta',\mathbf{r}')}{r(\theta)}
d\mu'(\theta',\mathbf{r}')>0.
\label{eq78gj}
\end{equation}
The last integral is strictly positive because 
$\mu'(A\cap I)$ is nonzero,
$r_\mathcal{C}(\theta,\theta',\mathbf{r}')$ is bounded away from
zero
by~(\ref{eq78ja}), and $r(\theta)$ is bounded above by~(\ref{eq92lk}). 
\end{Proof}

\medskip

\begin{Proof}{ of Theorem~\ref{t:ergodicity}}
First we will establish the $\psi$-irreducibility and 
aperiodicity \cite{meyn-tweedie:book2}
of the chain $\{S_i\}$. In fact, we will show that,
for any $n\geq 3$ and any state $s\in{\cal S}$, the measure
$\psi(\cdot)$ is absolutely continuous with respect to the measure
$P(S_{i+n}\in\cdot|S_i=s)$. To that end, choose and fix
an arbitrary state $s\in{\cal S}$ and an arbitrary
measurable subset $A$ of ${\cal S}$ with $\psi(A)>0$,
so that either
$({\cal L}_1\times\delta_\mathbf{0})(A)=\mu(A)>0$ or
$({\cal L}_1\times{\cal L}_2)(A)>0$ (or both).

In the first case, Lemma~\ref{lem:doeblin} implies
that $P(S_{i+2}\in A|S_i=s')>0$ for any $s'$, 
which, together with the Markov property, implies
that $P(S_{i+n}\in A|S_i=s)>0$ for all $n\geq 2$.
In the second case, 
combining Lemma~\ref{lem:doeblin} with
Lemma~\ref{lem:irred} applied to $A\cap{\cal S}_W$
and with the Markov
property, we obtain that there are
$\theta_1'<\theta_2'$ such that
$$
P(S_{i+3}\in A|S_i=s)
\geq
P(S_{i+3}\in A,S_{i+2}\in(\theta_1',\theta_2')\times\{\mathbf{0}\}|S_i=s)
\geq
\epsilon
\int_{\theta_1'}^{\theta_2'}
P(S_{i+3}\in A|S_{i+2}=(\theta,\mathbf{0}))\,d\theta,
$$
where the positivity of the last integral follows 
again from Lemma~\ref{lem:irred}. Finally, using the
Markov property once again, we have 
that $P(S_{i+n}\in A|S_i=s)>0$ for all $n\geq 3$,
as required.

Now, $\psi$-irreducibility and aperiodicity,
together with the Doeblin bound of Lemma~\ref{lem:doeblin},
imply \cite{meyn-tweedie:book2,kontoyiannis-meyn:I},
that the chain is uniformly ergodic. Specifically, 
Lemma~\ref{lem:doeblin} implies that the state space 
${\cal S}$ is small, and that the drift condition~(V4) 
of \cite{meyn-tweedie:book2} holds with Lyapunov 
function $V\equiv 1$. Then
\cite[Theorem~15.0.1]{meyn-tweedie:book2}
implies that the chain $\{S_i\}$ 
has a unique invariant (probability) measure $\pi$ 
to which the distribution of $S_i$ converges
uniformly, as stated in part~$1)$ of the theorem.
In particular, the chain $\{S_i\}$ is Harris recurrent,
and \cite[Theorem~17.0.1]{meyn-tweedie:book2}
implies that the strong law of large numbers holds 
for functions $F\in L_1(\pi)$,
as stated in part~$2)$ of the theorem.
\end{Proof}

\medskip

\begin{Proof}{ of Corollary~\ref{cor:ets}}
Since $X_{W,i}$ and $C_i=C(X_{W,i},Y_{W,i})$ are bounded,
and hence $\pi$-integrable,
functions of $S_i=(\Theta_i,(X_{W,i},Y_{W_i}))$,
the first two results immediately follow from 
Theorem~\ref{t:ergodicity}. For the next two,
let $\bar{\psi}$ denote the measure 
$\bar{\psi}=
{\cal L}_1\times\delta_\mathbf{0}\times[0,\infty)
+{\cal L}_1\times{\cal L}_2\times\delta_{0}$
on $\bar{\cal S}$.
Arguing as in 
the proof of Theorem~\ref{t:ergodicity},
it is easy to show that the new chain
$\{\bar{S}_i\}$ is $\bar{\psi}$-irreducible
and aperiodic, and also uniformly ergodic.
Once again, \cite[Theorem~17.0.1]{meyn-tweedie:book2}
implies that the strong law of large numbers holds 
for $\{\bar{S}_i\}$,
and recalling that $X_{B,i}=v_0\Delta_i\cos \Theta_i$,
the last two statements
of the corollary will follow as soon as we establish
that $\Delta$ is $\pi$-integrable. Indeed,
since, given $\Theta=\theta$, $\Delta$ is exponential with rate 
$r(\theta)\geq r_0>0$, we have
$$
E_{\bar{\pi}}(\Delta)
=E_\pi[E_{\bar{\pi}}(\Delta|\Theta)]
=E_\pi\Big[\frac{1}{r(\Theta)}\Big]\leq\frac{1}{r_0}<\infty,$$
completing the proof.
\end{Proof}

\subsection{The invariant distribution and expectations under $\pi$, $\bar{\pi}$}
\label{s:pi}

Let $\psi_B(\theta)$, $\theta\in[-\pi,\pi)$, be the density of
$\Theta_i\in{\cal S}_B$ under $\pi$, so that we think of
$\psi(\theta)d\theta$ is the long-term infinitesimal proportion
of time that $\Theta_i\in[\theta,\theta+d\theta) \subset \mathcal{S}_B$. 
Similarly, let $\psi_W(\theta,\mathbf{r})$, 
$(\theta,\mathbf{r})\in \mathcal{S}_W$,
denote the joint density of $(\Theta_i, (X_{W_i},Y_{W_i}))\in{\cal S}_W$
under $\pi$, so that we think of 
$\psi_W(\theta)d\theta dA$ as the long-term infinitesimal proportion
of time that $(\Theta_i,(X_{W,i},Y_{W,i}))\in
[\theta,\theta+d\theta)\times dA(\textbf{r}) \subset \mathcal{S}_W$,
where $dA(\textbf{r})$ is the infinitesimal area element $dA$ 
centered at $\textbf{r}$. 
In order to compute the functions
$\psi_B(\theta)$ and $\psi_W(\theta,\textbf{r})$, 
we derive balance equations, as follows. 

First, note that the proportion of state transitions out of 
$[\theta,\theta+d\theta]$ is $\psi_B(\theta)d \theta$. This should be equal 
to the proportion of transitions into $[\theta,\theta+d\theta]$,
for which we observe that the proportion of transitions out of 
$[\theta',\theta'+d\theta']$ is equal to $\psi_B(\theta')d\theta'$, 
and of these a proportion 
$\frac{r_\mathcal{A}(\theta',\theta)}{r(\theta')}d\theta$ transitions 
to states inside  $[\theta,\theta+d\theta]$. 
Likewise, the proportion of transitions out 
$[\theta,\theta+d\theta]\times dA(\textbf{r}')$, where $dA(\textbf{r}')$ is 
the infinitesimal area element centered at $\mathbf{r}'$ of size $dA'$, 
is $\psi_W(\theta,\mathbf{r}')d\theta dA'$. Of these, a proportion
$P_E(\theta,\mathbf{r}')$ enters $[\theta,\theta+d\theta]$. 
Therefore, 
integrating over all relevant state transitions we have
\begin{equation}
\psi_B(\theta) = \int_{-\pi}^\pi \psi_B (\theta') \frac{r_\mathcal{A}(\theta',\theta)}{r(\theta')}\, d\theta' + \int_\mathcal{F} \psi_W(\theta,\textbf{r}')P_E(\theta,\textbf{r}')\, dA'. 
\label{bal1}
\end{equation}

Second, observe that the proportion of transitions out of 
$[\theta,\theta+d\theta] \times dA(\textbf{r})$ is $\psi_W(\theta,\textbf{r})d\theta dA$. This should be equal to the proportion of transitions into that set. 
To compute that, we observe that the state is in $[\theta', \theta'+d\theta']$ 
for a proportion of time equal to $\psi_B(\theta')d\theta'$ and, 
at these instances, a proportion of transitions equal to
$$\frac{r_\mathcal{B}(\theta',\theta,\textbf{r})+
r_\mathcal{C}(\theta',\theta,\textbf{r})+
r_{\hat{\mathcal{D}}}(\theta',\theta,\textbf{r})}{r(\theta')}d \theta dA,$$
is into $[\theta,\theta+d\theta] \times dA(\textbf{r})$. Likewise,
the proportion of transitions out of 
$[\theta',\theta'+d\theta'] \times dA(\textbf{r}')$ 
is $\psi_W(\theta',\mathbf{r}')d\theta' dA'$. Of these, a proportion
$g(\theta,\mathbf{r};\theta',\mathbf{r}')d\theta dA$ enters 
$[\theta,\theta+d\theta] \times dA(\textbf{r})$. Again,
integrating over all relevant state transitions we obtain,
\begin{equation}
\psi_W(\theta,\textbf{r})= \int_{-\pi}^\pi \psi_B(\theta') 
\frac{r_\mathcal{B}(\theta',\theta,\textbf{r})+r_\mathcal{C}(\theta',\theta,\textbf{r})+r_{\hat{\mathcal{D}}}(\theta',\theta,\textbf{r})}{r(\theta')}\,d\theta' +\int_{-\pi}^\pi \int_F \psi_W(\theta',\textbf{r}')g(\theta,\textbf{r}; \theta',\textbf{r}')\,dA'd\theta'.
\label{bal2}
\end{equation}

In order to compute $\psi_B(\theta)$ and $\psi_W(\theta,\textbf{r})$, we discretize their arguments, which converts the balance equations (\ref{bal1}) and (\ref{bal2}) into a large linear system; we also use the fact that the sum of their integrals should be equal to unity; see Section \ref{s:linear} for details.

Then, having $\psi_B(\theta)$ and $\psi_W(\theta,\textbf{r})$, we can readily derive the following expectations with respect to the invariant distribution of the chain:
\begin{eqnarray*}
E_\pi(X_W)&=& \int_{-\pi}^{\pi} \int_\mathcal{F} \psi_W(\theta,\textbf{r})x\,dA d\theta, \\
E_\pi(C)&=& \int_{-\pi}^\pi \int_\mathcal{F} \psi_W(\theta,\textbf{r}) C(\textbf{r})\,dA d\theta,\\
E_{\bar{\pi}}(\Delta)&=&\int_{-\pi}^{\pi} \psi_B(\theta)\frac{1}{r_0(\theta)}\, d\theta, \\
E_{\bar{\pi}}(X_B)&=&\int_{-\pi}^\pi \psi_B(\theta) v_0\cos \theta \frac{1}{r_0(\theta)} \, d\theta,
\end{eqnarray*}
where the inner variable of integration in the first integral is $\textbf{r}=(x,y)$.

\subsection{Numerical computation of integrals}
In a number of instances in this work we need to compute the values
of a multivariate function that is given as a multiple integral. 
A good example is (\ref{eq38}), which is of the following form:
\begin{equation}
f(\theta,\mathbf{r})= \int_{-\pi}^\pi \iint_{\mathcal{F}}I(\theta,\mathbf{r};\theta',\mathbf{r}')\, dA'd\theta'. 
\label{eq38qq}
\end{equation}
We show how we calculate this integral; all other similar integrals in this work are calculated using the same method, mutatis mutandis. 

First, we discretize the variable $\theta$, considering only the $N$ values, 
\begin{equation}
\theta_i=-\pi+\frac{\pi}{N}(2i-1), \quad i=1,2,\ldots,N,
\label{eq6768}
\end{equation} 
where $N$ is a positive integer. 
Second, we discretize $\mathbf{r}$, by placing $M$ points uniformly inside 
$\mathcal{F}$. This is achieved as follows: we create $L^2$ points,
\begin{equation}
(x_{k_1},y_{k_2})=B\left( -1+\frac{2k_1-1}{L},~ -1+\frac{2k_2-1}{L}\right), 
\quad k_1,k_2=1,2,\dots,L,
\label{eq4537}
\end{equation}
where $L$ is a positive integer, and the parameter $B$ is such that the region $\mathcal{F}$ (which we have assumed bounded) lies entirely within the square $[-B,B] \times [-B,B]$; then, we keep those $M$ points that are within the region $\mathcal{F}$, denoting them as,
\begin{equation}
\mathbf{r}_j=(r_j,\phi_j)=(x_j,y_j),~j=1,2,\ldots M,
\label{eq67685}
\end{equation}
in polar and Cartesian coordinates respectively. 

Then, we restrict ourselves to calculating $f(\theta_i,\mathbf{r}_j)$ for 
all $1\leq i\leq N$, $1\leq j\leq M$, and 
equation~(\ref{eq38qq}) becomes,
\begin{equation}
f(\theta_i,\mathbf{r}_j)= \int_{-\pi}^\pi \iint_{\mathcal{F}}I(\theta_i,\mathbf{r}_j;\theta',\mathbf{r}') dA' d\theta'. 
\label{eq5gb}
\end{equation} 
This integral is calculated approximately as follows: We associate with each 
point $(\theta_{i'},\mathbf{r}_{j'})$, $1\leq i'\leq N$,
$1\leq j'\leq M$, the rectangle,
\begin{equation}
\left[\theta_{i'}-\frac{\pi}{N},\theta_{i'}+\frac{\pi}{N}\right] \times \left[x_{j'}-\frac{B}{L},x_{j'}+\frac{B}{L}\right] \times  \left[y_{j'}-\frac{B}{L},y_{j'}+\frac{B}{L}\right],
\label{eq88jj}
\end{equation}
of volume,
\begin{equation}
V_d=\frac{2\pi}{N} \times \frac{2B}{L} \times \frac{2B}{L} =\frac{8\pi B^2}{NL^2},
\label{eq7u21}
\end{equation} 
so that these rectangles approximately partition the set
$[-\pi,\pi]\times \mathcal{F}$ over which the integral (\ref{eq5gb}) is taken. 
And finally we set:
\begin{equation*}
f(\theta_i,\mathbf{r}_j) = V_d \sum_{i'=1}^N \sum_{j'=1}^M I(\theta_i,\mathbf{r}_j;\theta_{i'},\mathbf{r}_{j'}). 
\end{equation*} 
Clearly, the larger $N$ and $M$ are, the closer this approximation
to the true values of $f(\theta,\mathbf{r})$.

\subsection{Speeding up the computation of $E(N;\theta,\mathbf{r})$}
A time-consuming part of the computations required for our 
numerical results is the repeated calculation of 
$E(N;\theta,\mathbf{r})$ using (\ref{eq38}). Indeed, 
$E(N;\theta,\mathbf{r})$ is a function of two variables, and finding each 
of its values requires the evaluation of a multiple integral. 
However, assuming that the potential function $U(\theta,\mathbf{r})$ 
is only a function of $\theta$ makes this computation simpler. 
Indeed, in this case we have,
\begin{eqnarray*}
\lefteqn{\int_{-\pi}^\pi \iint_{\mathcal{F}} f_D(\theta') \mathbf{1} [U(\theta',\mathbf{r}')>U(\theta,\mathbf{0}),~\mathbf{r}' \in \mathcal{G}(\mathbf{r})]\, dA'd\theta'}\\
&=& \int_{-\pi}^\pi \iint_{\mathcal{F}}f_D(\theta') \mathbf{1} [U(\theta')>U(\theta)] \mathbf{1}[\mathbf{r}' \in \mathcal{G}(\mathbf{r})]\, dA'd\theta'\\
&=& \left( \int_{-\pi}^\pi f_D(\theta')\mathbf{1} [U(\theta')>U(\theta)] \, d\theta' \right) \left(\iint_{\mathcal{F}} \mathbf{1} [\mathbf{r}' \in \mathcal{G}(\mathbf{r})] \, dA'\right) = I_1(\theta) I_2(\mathbf{r}),
\end{eqnarray*}
where,
\begin{eqnarray}
I_1(\theta)  &=& \int_{-\pi}^\pi f_D(\theta')  \mathbf{1} [U(\theta')>U(\theta)]\,d\theta', \label{eq1001}\\
I_2(\mathbf{r}) &=& \iint_{\mathcal{F}} \mathbf{1}[\mathbf{r}' \in \mathcal{G}(\mathbf{r})]\, dA'. \label{eq1002}
\end{eqnarray}
Both of these integrals can be computed beforehand for all necessary
values of $\theta$ and $\mathbf{r}$, either numerically or analytically, 
and be readily available when the calculation of 
$E(N;\theta,\mathbf{r})$ starts. 

Furthermore, when $U(\theta,\mathbf{r})=-|\theta|$ then we have 
an even simpler expression for $I_1$:
\begin{equation*}
I_1(\theta) = \int_{-\pi}^\pi f_D(\theta') \mathbf{1} [|\theta'|<|\theta|]\,d\theta'=\int_{-|\theta|}^{|\theta|}f_D(\theta')\,d\theta'.
\end{equation*}

\subsection{Speeding up the computation of $g(\theta',\mathbf{r}';\theta,\mathbf{r})$} 

Another time-consuming part of our computations is the calculation 
of $g(\theta',\mathbf{r}';\theta,\mathbf{r})$, using (\ref{eq451d7}). 
This is because $g(\theta',\mathbf{r}';\theta,\mathbf{r})$ is a function of 
four variables, and each of its values requires the evaluation
of a multiple integral. The calculation can be significantly simplified if 
we take into account any special structure of the FR and the potential 
function. For example, if $U(\theta,\mathbf{r})$ is only a function 
of $\theta$, then (\ref{eq451d7}) 
simplifies to,
\begin{eqnarray*}
\lefteqn{\int_{-\pi}^\pi \iint_{\mathcal{F}} f_D(\theta'') \mathbf{1} [U(\theta'',\mathbf{r}'')>U(\theta',\mathbf{r}'),\mathbf{r}'' \in \mathcal{G}(\mathbf{r})]\, dA''d\theta''}\\
&=& \int_{-\pi}^\pi \iint_{\mathcal{F}}f_D(\theta'') \mathbf{1} [U(\theta'')>U(\theta')]\mathbf{1}[\mathbf{r}'' \in \mathcal{G}(\mathbf{r})]\, dA''d\theta''\\
&=& \left(\int_{-\pi}^\pi f_D(\theta'') \mathbf{1} [U(\theta'')>U(\theta')]\,d\theta''\right) \left(\iint_{\mathcal{F}}\mathbf{1}[\mathbf{r}'' \in \mathcal{G}(\mathbf{r})]\, dA''\right) = I_1(\theta') I_2(\mathbf{r}),
\end{eqnarray*}
where $I_1(\theta')$ and $I_2(\mathbf{r})$ are given by (\ref{eq1001}) and (\ref{eq1002}).

\subsection{Speeding up the computation of $r_\mathcal{B}(\theta,\theta',\mathbf{r}')$} 
One final computational bottleneck is the calculation of 
$r_\mathcal{B}(\theta,\theta',\mathbf{r}')$ using (\ref{rate_b}),
because it is a function of three variables and calculating any
of its values again involves the evaluation of a multiple integral. 
Considering, as before, the case where the
potential $U(\theta,\mathbf{r})$ is a function of $\theta$ only, 
the first integral in (\ref{rate_b}) becomes,
\begin{equation*}
\int_{-\pi}^{\pi} f_D(\theta'')\mathbf{1}\left[U(\theta',\mathbf{r}')>U(\theta'',\mathbf{0})\right]\, d\theta''= \int_{-\pi}^{\pi} f_D(\theta'') \mathbf{1} \left[ |\theta'|<|\theta''|\right]\, d\theta'' =I_3(\theta'),
\end{equation*}
where,
\begin{equation*}
I_3(\theta) = \int_{-\pi}^{\pi} f_D(\theta') \mathbf{1} \left[ |\theta|<|\theta'|\right]\, d\theta'.
\end{equation*}
Furthermore, the multiple integral in (\ref{rate_b}) becomes,
\begin{eqnarray*}
\lefteqn{\int_{-\pi}^{\pi} \iint_{\mathcal{F}} f_D(\theta''') \mathbf{1}[U(\theta,\mathbf{0})\geq U(\theta''',\mathbf{r}''')>U(\theta',\mathbf{r}')]\, dA'''d\theta'''} \\
&=& \int_{-\pi}^{\pi} \iint_{\mathcal{F}} f_D(\theta''') \mathbf{1}[|\theta|\leq|\theta'''|<|\theta'|]\, dA'''\,d\theta'''
= |\mathcal{F}| I_4(\theta,\theta'),
\end{eqnarray*}
where $|\mathcal{F}|$ is the area of $\mathcal{F}$, and,
\begin{equation*}
I_4(\theta,\theta')=\int_{-\pi}^\pi f_D(\theta'') \mathbf{1}[|\theta|\leq |\theta''|<|\theta'|] \, d\theta''.
\end{equation*}
In order to speed up the relevant calculations, 
$|\mathcal{F}|$, $I_3(\theta')$, and $I_4(\theta,\theta')$ can
be computed once in the beginning,
and made available when the calculation 
of $r_\mathcal{B}(\theta,\theta',\mathbf{r}')$ starts.

\subsection{Numerical solution of (\ref{bal1}) and (\ref{bal2})}
\label{s:linear}

Finally, here we describe the numerical solution of the system of 
balance equations~(\ref{bal1}) and~(\ref{bal2}) derived in 
Section~\ref{s:pi} of the Appendix above. As discussed there,
we restrict ourselves to computing
$\psi_B(\theta)$ and $\psi_W(\theta,\mathbf{r})$ only for the
discrete sets of values of $\theta$ and $\mathbf{r}$ 
given in~(\ref{eq6768}) and~(\ref{eq4537}), respectively,
so that~(\ref{bal1}) and~(\ref{bal2}) become,
\begin{eqnarray*}
\psi_B(\theta_i) &=& \int_{-\pi}^\pi \psi_B (\theta') \frac{r_\mathcal{A}(\theta',\theta_i)}{r(\theta')}\, d\theta' + \int_\mathcal{F} \psi_W(\theta_i,\textbf{r}')P_E(\theta_i,\textbf{r}')\, dA' \\
\psi_W(\theta_i,\textbf{r}_j)&=& \int_{-\pi}^\pi \psi_B(\theta') 
\frac{(r_\mathcal{B}+r_\mathcal{C}+r_{\hat{\mathcal{D}}})(\theta',\theta_i,\textbf{r}_j)}{r(\theta')}\,d\theta' +\int_{-\pi}^\pi \int_F \psi_W(\theta',\textbf{r}')g(\theta_i,\textbf{r}_j; \theta',\textbf{r}')\,dA'd\theta', 
\end{eqnarray*}
for $1\leq i \leq N$ and $1 \leq j \leq M$. Then we perform
a piecewise constant approximation of the integrands 
in the rectangular sets (\ref{eq88jj}), with the constant values 
being the values of the integrands at their centers. And taking
the integral over the union of these rectangles, we get,
\begin{eqnarray*}
\psi_B(\theta_i) &=& \sum_{i'} \psi_B (\theta_{i'}) \frac{r_\mathcal{A}(\theta_{i'},\theta_i)}{r(\theta_{i'})} \delta \theta + \sum_{j'} \psi_W(\theta_i,\textbf{r}_{j'})P_E(\theta_i,\textbf{r}_{j'})  \delta A, \\
\psi_W(\theta_i,\textbf{r}_j)&=& \sum_{i'}\psi_B(\theta_{i'}) 
\frac{(r_\mathcal{B}+r_\mathcal{C}+
r_{\hat{\mathcal{D}}})(\theta_{i'},\theta_i,\textbf{r}_j)}{r(\theta_{i'})}\delta \theta  +\sum_{i',j'}\psi_W(\theta_{i'},\textbf{r}_{j'})g(\theta_i,\textbf{r}_j; \theta_{i'},\textbf{r}_{j'}) \delta A \delta \theta, 
\end{eqnarray*}
where $\delta \theta = \frac{2\pi}{N}$ and 
$\delta A=\left(\frac{2B}{L}\right)^2$. Multiplying
the first equation by $\delta \theta$, the second equation 
by $\delta \theta \delta A$, and defining, with a slight abuse of notation, 
\begin{equation*}
\psi_B(i)\triangleq \psi_B(\theta_i)\delta \theta, \quad \psi_W(i,j) \triangleq \psi(\theta_i,\mathbf{r}_j),
\end{equation*}
and likewise for all other functions appearing in the balance equations,
we obtain the system,
\begin{eqnarray*}
\psi_B(i) &=& \sum_{i'} \psi_B (i') \frac{r_\mathcal{A}(i',i)}{r(i')} \delta \theta + \sum_{j'} \psi_W(i,j')P_E(i,j'),  \\
\psi_W(i,j)&=& \sum_{i'}\psi_B(i') \frac{(r_\mathcal{B}+r_\mathcal{C}+r_{\hat{\mathcal{D}}})(i',i,j)}{r(i')}\delta \theta \delta A  +\sum_{i',j'}\psi_W(i',j')g(i,j; i',j') \delta A \delta \theta,
\end{eqnarray*}
for $1\leq i \leq N$ and $1 \leq j \leq M$. 
This is a linear system of $N+MN=N(M+1)$ equations,
which may be 
interpreted as expressing the balance equations of a discrete 
Markov chain with $N(M+1)$ states (because (\ref{eq4ho}) and (\ref{eq4ho9})
ensure that the probabilities of transitions out of each state indeed 
add up to unity in all cases). To solve this system, we write it in the 
form $\psi=K \psi$, where $\psi$ is a vector of length $N(M+1)$ consisting
of $\psi_B$ followed by each of the columns of $\psi_W$.
Since the resulting matrix $K$ is stochastic, by construction, we find 
an eigenvector corresponding to its top eigenvalue, which, after
normalization, provides the required solution.


\end{document}